\documentclass[aps,twocolumn,pra,superscriptaddress,floatfix,showpacs,tightenlines]{revtex4}
\usepackage{epsfig,graphicx,times}
\usepackage{amstext}
\usepackage{amsmath}
\usepackage{amssymb}
\usepackage{graphicx}
\usepackage{latexsym}
\usepackage{bm}

\def\dbarit {{\mathchar'26\mkern-11mud}}
\input{tcilatex}

\begin{document}

\title{Quantum Thermodynamic Cycles and Quantum Heat Engines}
\author{H.T. Quan}
\affiliation{Frontier Research System, The Institute of Physical and Chemical Research
(RIKEN), Wako-shi, Saitama 351-0198, Japan}
\affiliation{Institute of Theoretical Physics, The Chinese Academy of Sciences, Beijing,
100080, China}
\author{ Yu-xi Liu}
\affiliation{CREST, Japan Science and Technology Agency (JST), Kawaguchi, Saitama
332-0012, Japan}
\affiliation{Frontier Research System, The Institute of Physical and Chemical Research
(RIKEN), Wako-shi, Saitama 351-0198, Japan}
\author{C. P. Sun}
\affiliation{Frontier Research System, The Institute of Physical and Chemical Research
(RIKEN), Wako-shi, Saitama 351-0198, Japan}
\affiliation{Institute of Theoretical Physics, The Chinese Academy of Sciences, Beijing,
100080, China}
\author{Franco Nori}
\affiliation{Frontier Research System, The Institute of Physical and Chemical Research
(RIKEN), Wako-shi, Saitama 351-0198, Japan}
\affiliation{Center for Theoretical Physics, Physics Department, Center for the Study of
Complex Systems, The University of Michigan, Ann Arbor, Michigan 48109-1040,
USA}
\date{\today }

\begin{abstract}
In order to describe quantum heat engines, here we systematically study
isothermal and isochoric processes for quantum thermodynamic cycles. Based
on these results the quantum versions of both the Carnot heat engine and the
Otto heat engine are defined without ambiguities. We also study the
properties of quantum Carnot and Otto heat engines in comparison with their
classical counterparts. Relations and mappings between these two quantum
heat engines are also investigated by considering their respective quantum
thermodynamic processes. In addition, we discuss the role of Maxwell's demon
in quantum thermodynamic cycles. We find that there is no violation of the
second law, even in the existence of such a demon, when the demon is
included correctly as part of the working substance of the heat engine.
\end{abstract}

\maketitle

\pagenumbering{arabic}

\section{INTRODUCTION}

Quantum heat engines (QHEs) \cite{scovil1,scovil2} produce work using
quantum matter as their working substance. Because of the quantum nature of
the working substance, QHE have unusual and exotic properties. For example,
under some conditions, QHE can surpass the maximum limit on the amount of
work done by a classical thermodynamic cycle \cite{kieu1,kieu2} and also
surpass the efficiency of a classical Carnot engine cycle \cite%
{scully-science}. QHEs offer good model systems to study the relation
between thermodynamics and quantum mechanics. Meanwhile, they can highlight
the difference between classical and quantum thermodynamic systems, and help
us understand the quantum-classical transition problem of thermodynamic
processes \cite{quan2}.

The classical Carnot heat engine is a well-known machine that produces work
through thermodynamic cycles. The thermodynamic properties of the four
strokes of each cycle are simple and demonstrate the universal physical
mechanism of heat engines. Current studies \cite%
{kieu1,kieu2,scully-science,quan2,bender1,bender2,scully-pra,opatrny,pce,arnaud1,arnaud2,mahler}
on QHE mostly focus on the quantum analogue of classical Carnot engines,
i.e., the quantum Carnot engine (QCE). The quantum Otto engine (QOE) is
another interesting case of a QHE, which is also attracting considerable
attentions \cite{kieu1,kieu2,kosloff1,kosloff2,quan1,quan4}. However, we
find that there is no universal and consistent definition of the QCE and the
QOE in literatures (see, e.g., Refs. \cite%
{scully-science,opatrny,scully-pra,scully2,allahverdyan}), and\ thus the
properties of QCEs and QOE are not always addressed adequately and clearly.

Any QHE cycle consists of several basic quantum thermodynamic processes,
such as quantum adiabatic processes (which have been clarified in many
references, e.g., \cite{quantum adiabatic}), quantum isothermal processes
and quantum isochoric processes. This paper begins by clarifying the
concepts of isothermal processes, isochoric processes and effective
temperatures in their quantum mechanical pictures. Then we systematically
study the general properties of a quantum analogue of a Carnot engine. The
difference between a QCE and its classical counterpart is indicated clearly.
We also study the QOE based on its basic quantum thermodynamic process and
analyze the relation between these two types of QHEs. Here we assume that
the processes of our thermodynamic cycles are infinitely slow, i.e., the
time interval of each process is assumed to be very long. Accordingly, the
output power is very small. This is also the requirement of quasi-static
processes. Assuming fast cycles would increase the output powers, but at the
expense of reduced engine efficiency. Also, some experimentally realizable
physical systems, which can be used to implement our QCE and QOE, are
discussed. Furthermore, based on our generalized QOE model, we demonstrate
that there is no violation of the second law, even in the presence of a
Maxwell's demon.

Our paper is organized as follows: In Sec.~II we give a clear definition of
quantum isothermal and isochoric processes based on the quantum
identification of work performed and heat exchange. In Sec.~III we discuss
the QCE cycle and calculate the work done during this cycle and its
operation efficiency. In Sec.~IV we discuss the QOE cycle and compare it
with the classical Otto engine cycle. In Sec.~V we compare these two kinds
of QHEs and study the relation between them. In Sec.~VI we give some
examples of these two kinds of QHEs considering experimentally-realizable
physical systems. In Sec.~VII we discuss the QOE and Maxwell's demon.
Conclusions and remarks are given in Sec.~VIII.

\section{BASIC QUANTUM THERMODYNAMIC PROCESS}

\subsection{Quantum first law of thermodynamics}

To define quantum isothermal and quantum isochoric processes, we need to
first consider the working substance. An arbitrary quantum system with a
finite number of energy levels is used here as the working substance (see
Fig.~1). (Of course, this can be generalized to systems with an infinite
number of energy levels). The Hamiltonian of the working substance can be
written as%
\begin{equation}
H=\sum_{n}E_{n}\left\vert n\right\rangle \left\langle n\right\vert ,
\label{1}
\end{equation}%
where $\left\vert n\right\rangle $ is the $n$-th eigen state of the system
and $E_{n}$ is its corresponding eigen energy. Without loss of generality,
we choose the eigen energy of the ground state $\left\vert 0\right\rangle $
as a reference point (see Appendix A). Then the Hamiltonian (\ref{1}) can be
rewritten as%
\begin{equation}
H=\sum_{n}(E_{n}-E_{0})\left\vert n\right\rangle \left\langle n\right\vert .
\label{2}
\end{equation}%
Below we will show that it is convenient for our discussion about QHEs to
use the Hamiltonian (\ref{2}). The internal energy $U$ of the working
substance can be expressed as%
\begin{equation}
U=\left\langle H\right\rangle =\sum_{n}P_{n}E_{n},  \label{1.5}
\end{equation}%
for a given occupation distribution with probabilities $P_{n}$ in the $n$-th
eigen state.

To clearly define quantum isothermal and isochoric processes, we need to
identify the quantum analogues of the heat exchange $\dbarit Q$ and the work
performed $\dbarit W$.\ From Eq.~(\ref{1.5}) we have%
\begin{equation}
dU=\sum_{n}\left[ E_{n}\,dP_{n}+P_{n}\,dE_{n}\right] .  \label{2.1}
\end{equation}%
In classical thermodynamics, the first law of thermodynamics is expressed as 
\begin{equation}
dU=\dbarit Q+\dbarit W,  \label{2.5}
\end{equation}%
where $\dbarit Q=TdS$, and $\dbarit W=\sum_{i}Y_{i}dy_{i}$ \cite{callen}; $T$
is the temperature and $S$ is the entropy; $y_{i}$ is the generalized
coordinates and $Y_{i}$ is the generalized force conjugated to $y_{i}$. Due
to the relationship $S=-k_{B}\sum_{i}P_{i}\ln P_{i}$ between the entropy $S$
and the probabilities $P_{i}$, we can make the following identification \cite%
{kieu1,kieu2,quan1}%
\begin{eqnarray}
\dbarit Q &=&\sum_{n}E_{n}\,dP_{n},  \label{3} \\
\dbarit W &=&\sum_{n}P_{n}\,dE_{n}.  \label{4}
\end{eqnarray}%
Equation (\ref{4}) implies that the work performed corresponds to the change
in the eigen energies $E_{n}$, and this is in accordance with the fact that
work can only be performed through a change in the generalized coordinates
of the system, which in turn gives rise to a change in the eigen energies 
\cite{schrodinger, kieu2}. Thus the quantum version of the first law of
thermodynamics $dU=\dbarit Q+\dbarit W$ just follows from Eq.~(\ref{2.1})
with the quantum identifications of heat exchange and work performed in
Eqs.~(\ref{3}) and (\ref{4}). Different from $\dbarit Q=TdS$, which is
applicable only to the thermal equilibrium case, below we will see that
Eqs.~(\ref{3}) and (\ref{4}) are applicable to both the thermal equilibrium
case (see e.g., Eq. (\ref{24})) and the nonequilibrium case (see e.g., Eq. (%
\ref{11})).

%Figure 1
%\begin{figure}[h]
\begin{figure}[tbp]
\begin{center}
\includegraphics[bb=29 331 572 810, width=8cm, clip]{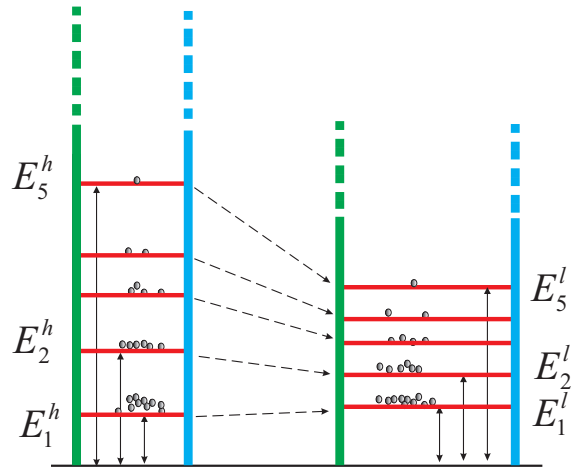}
\end{center}
\caption{(Color online) Schematic diagram of multi-level quantum system as
the working substance for a QHE. $E_{n}^{h}$ and $E_{n}^{l}$ are the $n$-th
eigen energy of the working substance in the two isochoric processes.}
\end{figure}
\begin{table*}[tbp]
\caption{Quantum versus classical thermodynamic processes. Here we use
\textquotedblleft INV" to indicate the invariance of a thermodynamic
quantity and \textquotedblleft VAR" to indicate that it varies or changes. $%
U $ is the internal energy of the working substance; $T$, $P$, $E_{n}$, $%
P_{n}$ are defined in Sec. II. The working substance of the classical
thermodynamic processes considered here is the ideal classical gas. }
\begin{center}
\begin{tabular}{c|c|c|c}
\hline\hline
& isothermal process & isochoric process & adiabatic process \\ \hline\hline
\parbox{2.5cm} {Classical} & 
\parbox{5cm} {\ \ \ \ \ \ \ \
\ \ \ \ \ \ \ \ \ \ \ \ \ \ \ \ \ \ \ \ \ \ \ \ \ \ \ \ \ \ \ \ \ \
\ \ \ \ \ \ \ \ \ \ \ \ \ \ \ \ \ \ \ \ \ \ \ \ \ \ \ \ \ \ Heat
absorbed or released.\ \ \ \ \ \ \ \ \ \ \ \ \ \ \ \ \ \ \ \ Work
done.\ \ \ \ \ \ \ \ \ \ \ \ \ \ \ \ \ \ \ \ \ \ \ \ \ \ \ \ \ \ \ \
\ \ \ \ \ \ \ \ \ \ \ \ \ INV: \ \ $U$, $T$ \ \ \ \ VAR: \ \ $P$,
$V$ \ \ \ \ \ \ \ \ \ \ \ \ \ \ \ \ \ \ \ \ \ \ \ \ \ \ \ \ \ \ \ \
\ \ \ \ \ \ } & 
\parbox{4.5 cm} {Heat absorbed or released. \ \ \ \ \ \ \ \ \ \ \ \ \ No work done. \ \ \ \ \ \ \ \ \ \ \ \ \
\ \ \ \ \ \ \ \ \ \ \ \ \ \ \ \ \ \ \ \ \ \ \ \ \ INV: \ \ $V$  \ \
\ \ VAR: \ \ $P$, $T$} & 
\parbox{4 cm}{No heat exchange. \ \ \ \ \ \ \ \ \ \ \ \ \ \ \ \ \ \ \ \ \ \ \ \ \ \ \ \
\ \ \ \ \ \ Work done.  \ \ \ \ \ \ \ \ \ \ \ \ \ \ \ \ \ \ \ \ \ \
\ \ \ \ \ \ \ \ \ \ \ \ VAR: \ \ $P$, $T$, $V$} \\ \hline
\parbox{2.5cm} {Quantum} & 
\parbox{5cm} {\ \ \ \ \ \ \ \ \ \ \ \ \ \ \ \ \ \ \ \ \ \ \ \ \ \
\ \ \ \ \ \ \ \ \ \ \ \ \ \ \ \ \ \ \ \ \ \ \ \ \ \ \ \ \ \ \ \ \ \
\ \ \ \ \ \ \ \ \ \ \ \ Heat absorbed or released. \ \ \ \ \ \ \ \ \
\ \ \ \ \ \ \ \ \ \ \ Work done. \ \ \ \ \ \ \ \ \ \ \ \ \ \ \ \ \ \
\ \ \ \ \ \ \ \ \ \ \ \ \ \ \ \ \ \ \ \ \ \ \ \ \ INV: \ \ $T$ \ \ \
\ VAR: \ \ $U$, $E_{n}$, $P_{n}$ \ \ \ \ \ \ \ \ \ \ \ \ \ \ \ \ \ \
\ \ \ \ \ \ \ \ \ \ \ \ \ \ \ \ \ \ \ \ } & 
\parbox{4.5 cm} {Heat absorbed or released. \ \ \ \ \ \ \ \ \ \ \ \ \ No work done. \ \ \ \ \ \ \ \ \ \ \ \ \
\ \ \ \ \ \ \ \ \ \ \ \ \ \ \ \ \ \ \ \ \ \ \ \ \ \ \ \ \ \ \ \ \ \
\ \ \ \ \ \ \ \ \ \ \ \ \ \ \ \ \ \ \ \ \ \ \ \ \ \ \ \ \ \ \ INV: \
\ $E_{n}$ \ \ \ \ VAR: \ \ $P_{n}$, $T_{\mathrm{eff}}$} & 
\parbox{4 cm} {No heat
exchange. \ \ \ \ \ \ \ \ \ \ \ \ \ \ \ \ \ \ \ \ \ \ \ \ \ \ \ \ \
\ \ \ \ \ Work done. \ \ \ \ \ \ \ \ \ \ \ \ \ \ \ \ \ \ \ \ \ \ \ \
\ \ \ \ \ \ \ \ \ \ INV: \ \ $P_{n}$  \ \ \ \ VAR: \ \ $E_{n}$,
$T_{\mathrm{eff}}$} \\ \hline\hline
\end{tabular}%
\end{center}
\end{table*}

\subsection{Quantum isothermal process}

Let us now consider the quantum versions of some thermodynamic processes.
First we study quantum isothermal processes. In quantum isothermal
processes, the working substance, such as a particle confined in a potential
energy well, is kept in contact with a heat bath at a constant temperature.
The particle can perform positive work to the outside, and meanwhile absorb
heat from the bath. Both the energy gaps and the occupation probabilities
need to change simultaneously, so that the system remains in an equilibrium
state with the heat bath at every instant. Specifically, let us consider a
two-level system with the excited state $\left\vert e\right\rangle $, the
ground state $\left\vert g\right\rangle $, and a single energy spacing $%
\Delta $. In the quasi-static quantum isothermal process, the ratio $%
r=P_{e}/P_{g}$ of the two occupation probabilities, $P_{e}$ and $P_{g}$,
must satisfy the Boltzmann distribution $r=P_{e}/P_{g}=\exp [-\beta \Delta
(t)]$ and also the normalization condition $P_{e}+P_{g}=1$. $\Delta (t)$
changes slowly with time $t$, and accordingly $r$ can be written as 
\begin{equation}
r\equiv r(t)=\frac{P_{e}}{P_{g}}=e^{-\beta \Delta (t)},  \label{4.5}
\end{equation}%
where $\beta =1/k_{B}T$, $k_{B}$ is the Boltzmann constant and $T$ is the
temperature. In a sufficiently slow process, at every instant the system
remains in thermodynamic equilibrium with the heat bath.

\subsection{Effective temperature}

We can also define an effective temperature $T_{\mathrm{eff}}$ for any
two-level system according to the ratio $r(t)$ and the level spacing $\Delta
(t)$. For a two-level system with energy levels $E_{e}$ and $E_{g}$, even in
a non equilibrium state, we can imagine that it is in a virtual equilibrium
state with the effective temperature%
\begin{equation}
T_{\mathrm{eff}}=\frac{1}{k_{B}\beta _{\mathrm{eff}}}=\frac{\Delta (t)}{k_{B}%
}\left[ \ln \frac{P_{g}}{P_{e}}\right] ^{-1},  \label{4.2}
\end{equation}%
as long as the level spacing $\Delta (t)$ and the energy level distributions 
$P_{g}$ and $P_{e}$ are known. Of course, Eq.~(\ref{4.2}) cannot be directly
generalized to the case with more than two levels. For example, for a
three-level system with occupation probabilities $P_{a}$, $P_{b}$, and $%
P_{c} $ in three states denoted by $\left\vert a\right\rangle $, $\left\vert
b\right\rangle $, and $\left\vert c\right\rangle $, if the two level
spacings $\Delta _{ab}(t)$ and $\Delta _{bc}(t)$ do not satisfy the relation%
\begin{equation}
\frac{1}{\Delta _{ab}(t)}\ln \frac{P_{a}}{P_{b}}=\frac{1}{\Delta _{bc}(t)}%
\ln \frac{P_{b}}{P_{c}},  \label{4.6}
\end{equation}%
we cannot define a unique effective temperature. The subset $\{\left\vert
a\right\rangle $, $\left\vert b\right\rangle \}$ can have an effective
temperature defined by Eq.~(\ref{4.2}), while the subset $\{\left\vert
b\right\rangle $, $\left\vert c\right\rangle \}$ would have a different
effective temperature. We will discuss this point in detail in a QCE cycle
in Sec. III.

\subsection{Quantum isochoric process}

A quantum isochoric process has similar properties to that of a classical
isochoric processes. In a quantum isochoric process, the working substance
is placed in contact with a heat bath. No work is done in this process while
heat is exchanged between the working substance and the heat bath. This is
the same as that in a classical isothermal process. In a quantum isochoric
process the occupation probabilities $P_{n}$ and thus the entropy $S$\
change, until the working substance finally reaches thermal equilibrium with
the heat bath. In classical isochoric process the pressure $P$ and the
temperature $T$ change, and the working substance reaches thermal
equilibrium with the heat bath only at the end of this process. For example,
if the working substance is chosen to be a particle confined in a infinite
square well potential, no work is done during a quantum isochoric process
when heat is absorbed or released, and the occupation probabilities in every
eigen state satisfy Boltzmann distribution at the end of the isochoric
process.

\subsection{Quantum adiabatic process}

A classical adiabatic thermodynamic process can be formulated in terms of a
microscopic quantum adiabatic thermodynamic process. Because quantum
adiabatic processes proceed slow enough such that the generic quantum
adiabatic condition is satisfied, then the population distributions remain
unchanged, $dP_{n}=0$. According to Eq.~(\ref{3}), $\dbarit Q=0$, there is
no heat exchange in a quantum adiabatic process, but work can still be
nonzero according to Eq.~(\ref{4}). A classical adiabatic process, however,
does not necessarily require the occupation probabilities to be kept
invariant. For example, when the process proceeds very fast, and the quantum
adiabatic condition is not satisfied, internal excitations will likely
occur, but there is no heat exchange between the working substance and the
external heat bath. This thermodynamic process is classical adiabatic but
not quantum adiabatic. Thus it can be verified that a classical adiabatic
process includes, as a subset, a quantum adiabatic process; but the inverse
is not valid \cite{quan1}.

The properties of both classical and quantum thermodynamic process are
listed in Table I, to facilitate the comparison between these two processes.
The table indicates if heat is absorbed or released (first row), if work is
done (second row), and which quantity varies (indicated by \textquotedblleft
VAR") and which are invariant (indicated by \textquotedblleft INV") in the
third row. %\begin{widetext}

%\end{widetext}
%Figure 2
%\begin{figure}[h]
\begin{figure}[tbp]
\begin{center}
\includegraphics[bb=166 199 404 726, width=7.5cm, clip]{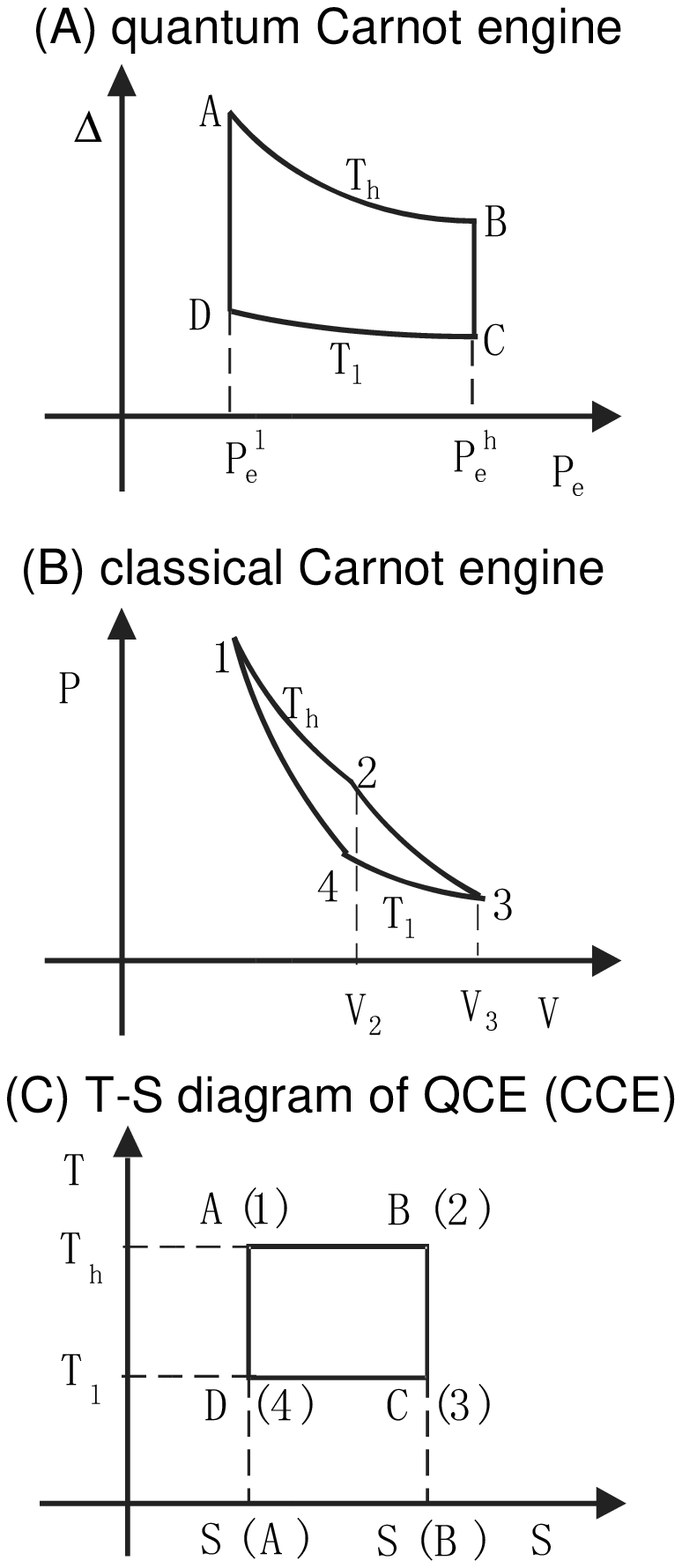}
\end{center}
\caption{(A): A schematic diagram of a \textit{quantum} Carnot engine based
on a two-level quantum system. $\Delta $ is the level spacing between the
two energy levels. $P_{e}$ is the occupation probability in the excited
state. The process from $A$ to $B$ ($C$ to $D$) is the isothermal expansion
(compression) process, in which the working substance is put in contact with
the high (low) temperature heat bath. The processes from $B$ to $C$ and from 
$D$ to $A$ are two adiabatic processes. (B): Pressure-Volume ($PV$) diagram
for a \textit{classical} Carnot engine with ideal gas as the working
substance. The process from $1$ to $2$ (from $3$ to $4$) is the classical
isothermal expansion (compression) process with temperature $T_{h}$ ($T_{l}$%
), and the process from $2$ to $3$ ($4$ to $1$) is the classical adiabatic
expansion (compression) process. $V_{2}$ and $V_{3}$ are the volume of the
working substance at $2$ and $3$ respectively. (C): Temperature-Entropy ($%
T-S $) diagram \protect\cite{schroeder} for both for a \textit{quantum}
Carnot engine based on a two-level quantum system and a \textit{classical}
Carnot engine with ideal gas as the working substance. This $T-S$ diagram
bridges the \textit{quantum} and \textit{classical} Carnot engine.}
\end{figure}

\section{QUANTUM CARNOT ENGINE CYCLE}

In the previous section, we defined quantum isothermal processes. Based on
this definition, in this section, we study the QCE cycle and its properties.
The QCE cycle (see Fig.~2 for an example of a QCE based on a two-level
system), just like its classical counterpart, consists of two quantum
isothermal processes ($A\longrightarrow B$ and $C\longrightarrow D$) and two
quantum adiabatic processes ($B\longrightarrow C$ and $D\longrightarrow A$).
During the isothermal expansion process from $A$ to $B$, the particle
confined in the potential well is kept in contact with a heat bath at
temperature $T_{h}$, while the energy levels of the system change much
slower than the relaxation of the system, so that the particle is always
kept in thermal equilibrium with the heat bath. Below, we consider both
cases: two-level and multi-level systems.

\subsection{Thermodynamic reversibility of the quantum Carnot engine cycle}

It is well known that quantum mechanical reversibility are associated with
quantum mechanical unitary evolution. Different from quantum mechanical
reversibility, thermodynamic reversibility accompanies the heat bath and the
effective temperature of the working substance. In this paper we focus on
the thermodynamic reversibility.

We emphasize that, in order to ensure that the cycle is thermodynamically 
\textit{reversible}, two conditions on the quantum adiabatic process are
required: (1) after the quantum adiabatic process ($B\longrightarrow C$), we
can use an effective temperature $T_{l}$ to characterize the working
substance, i.e., the working substance still satisfies the Boltzmann
distribution after the quantum adiabatic process; and (2) the effective
temperature $T_{l}$ of the working substance, after the quantum adiabatic
process, equals the temperature $T_{l}$ of the heat bath of the following
quantum isothermal process ($C\longrightarrow D$). When either condition is
not satisfied, a thermalization process \cite{orszag,thermolization} of the
working substance is inevitable before the quantum isothermal process ($%
C\longrightarrow D$). In the thermalization process, the total entropy
increase of the working substance plus the bath is nonzero. Hence, this
thermalization process is irreversible.

It can be proved that the above two conditions are equivalent to the
following two conditions: (i)\ all energy gaps were changed by the same
ratio in the quantum adiabatic process, i.e., $E_{n}(B)-E_{m}(B)=\lambda
\lbrack E_{n}(C)-E_{m}(C)]$,\ and $E_{n}(A)-E_{m}(A)=\lambda \lbrack
E_{n}(D)-E_{n}(D)]$, $(n=0,1,2,\cdots )$; and (ii) the ratio of the change
of the energy gaps in the adiabatic process (from $B$ to $C$ or from $A$ to $%
D$) must equal the ratio of the two temperatures of the heat baths, i.e., $%
\lambda =T_{h}/T_{l}$.

First, it is easy to see that these two conditions, (i) and (ii), listed
right above are sufficient for the previous conditions (1) and (2) presented
initially. Next, we prove that the two conditions (i) and (ii) are also
necessary for the two conditions (1) and (2). Let us assume that the working
substance is in equilibrium with a heat bath at temperature $T_{h}$ at the
instant $B$ before the adiabatic process ($B\longrightarrow C$). In this
case the quantum state is described by a density operator%
\begin{equation}
\rho (B)=\frac{1}{Z}\sum_{n}\exp [-\beta _{h}E_{n}(B)]\left\vert
n(B)\right\rangle \left\langle n(B)\right\vert .  \label{4.7}
\end{equation}%
After the adiabatic process is completed, at instant $C$ in Fig.~2, the
eigen energies of the working substance become $E_{n}(C)$, and the working
substance reaches an effective temperature $T_{l}$. The occupation
probabilities $P_{n}$ of the working substance are kept unchanged during the
adiabatic process ($B\longrightarrow C$), and they satisfy the Boltzmann
distribution. Thus, for any eigen states $\left\vert n\right\rangle $ and $%
\left\vert m\right\rangle $, the occupation probabilities $P_{n}$ and $P_{m}$
satisfy 
\begin{eqnarray}
\frac{P_{n}(B)}{P_{m}(B)} &=&\frac{\exp [-\beta _{h}E_{n}(B)]}{\exp [-\beta
_{h}E_{m}(B)]}  \notag \\
&=&\frac{P_{n}(C)}{P_{m}(C)}=\frac{\exp [-\beta _{l}E_{n}(C)]}{\exp [-\beta
_{l}E_{m}(C)]}.  \label{9.7}
\end{eqnarray}%
That is, 
\begin{equation}
E_{n}(C)-E_{m}(C)=\frac{T_{l}}{T_{h}}[E_{n}(B)-E_{m}(B)],  \label{10.1}
\end{equation}%
for any $m,n$. Eq. (\ref{10.1}) is just a combination of conditions (i) and
(ii). Thus we have proved that the two conditions (i) and (ii) are
sufficient for the previous conditions (1) and (2).

Hence, we have proven that (i)\ all energy gaps change by the same ratio in
quantum adiabatic process, and (ii) this ratio equals to the ratio of the
temperatures of the two heat baths, summarized in Eq.~(\ref{10.1}), are
sufficient and necessary conditions for the QCE to be thermodynamically
reversible. We would like to mention that these two conditions (i)\ and (ii)
(mathematical requirements) can be satisfied in some realistic physical
systems. Examples of QCEs based on some concrete physical systems will be
discussed in Sec. VI. 
\begin{table*}[tbp]
\caption{Quantum Carnot engines versus classical Carnot engines. Here
\textquotedblleft CIT" refers to \textquotedblleft Classical Iso-Thermal
process" while \textquotedblleft CA" is an abbreviation for a
\textquotedblleft Classical Adiabatic process". \textquotedblleft QIT" and
\textquotedblleft QA" refer to \textquotedblleft Quantum Iso-Thermal
process" and \textquotedblleft Quantum Adiabatic process", respectively. $%
V_{2}$, $V_{3}$, $E_{n}(B)$, and $E_{n}(C)$ are defined in Fig. 2; $\protect%
\gamma $ is the adiabatic exponent \protect\cite{schroeder}.}
\begin{center}
\begin{tabular}{c|c|c|c|c}
\hline\hline
& \parbox{3cm} {strokes} & \parbox{5cm} {requirement on
the CA and QA} & \parbox{3cm} {efficiency} & 
\parbox{4cm}{positive-work
condition} \\ \hline\hline
Classical & 
\parbox{3cm} { \ \ \ \ \ \ \ \ \ \ \ \ \
\ \ \ \ \ \ \ \ \ \ \ \ \ \ \ \ \ \ \ \ \ \ \ \ \ \ \ \ \ \ \ \ \ \
\ \ \ \ \ \ \ \ \ \ \ \ \ \ \ \ \ \ \ \ \ \ \ \ \ \ \ \ \ \ \ \ \ \
\ \ \ \ \ \ \ \ \ \ \ \ \ \ \ \ \ \ \ \ CIT-CA-CIT-CA \ \ \ \ \ \ \
\ \ \ \ \ \ \ \ \ \ \ \ \ \ \ \ \ \ \ \ \ \ \ \ \ \ \ \ \ } & 
\parbox{5cm}
{$\frac{T_{l}}{T_{h}}=[\frac{V_{2}}{V_{3}}]^{\gamma -1}$} & $\eta =1-\frac{%
T_{l}}{T_{h}}$ & $T_{h}>T_{l}$ \\ \hline
Quantum & 
\parbox{3cm} { \ \ \
\ \ \ \ \ \ \ \ \ \ \ \ \ \ \ \ \ \ \ \ \ \ \ \ \ \ \ \ \ \ \ \ \ \
\ \ \ \ \ \ \ \ \ \ \ \ \ \ \ \ \ \ \ \ \ \ \ \ \ \ \ \ \ \ \ \ \ \
\ \ \ \ \ \ \ \ \ \ \ \ \ \ \ \ \ \ \ \ \ \ \ \ \ \ \ \ \ \
QIT-QA-QIT-QA \ \ \ \ \ \ \ \ \ \ \ \ \ \ \ \ \ \ \ \ \ \ \ \ \ \ \
\ \ \ \ \ \ \ \ \ } & 
\parbox{5cm}{$\frac{T_{l}}{T_{h}}=\frac{E_{n}(C)-E_{m}(C)}{E_{n}(B)-E_{m}(B)}$ for
$\forall $ $m$, $n$} & $\eta =1-\frac{T_{l}}{T_{h}}$ & $T_{h}>T_{l}$ \\ 
\hline\hline
\end{tabular}%
\end{center}
\end{table*}

\subsection{Work and efficiency of a quantum Carnot engine cycle}

Now we analyze the operation efficiency $\eta _{\mathrm{C}}$ of the QCE
introduced above. For simplicity, instead of applying Eq.~(\ref{3}), we use $%
\dbarit Q=TdS$ to calculate the heat exchange $\dbarit Q$ in any Quantum
Iso-Thermal (QIT) process. Because the temperature of the heat bath is kept
invariant in the quantum isothermal process, the heat absorbed $Q_{\mathrm{in%
}}^{\mathrm{QIT}}$ and released $Q_{\mathrm{out}}^{\mathrm{QIT}}$ in the
quantum isothermal expansion and compression processes can be calculated as
follows%
\begin{eqnarray}
Q_{\mathrm{in}}^{\mathrm{QIT}} &=&T_{h}[S(B)-S(A)]>0,  \label{9.3} \\
Q_{\mathrm{out}}^{\mathrm{QIT}} &=&T_{l}[S(C)-S(D)]>0,  \label{9.6}
\end{eqnarray}%
where $T_{h}$ and $T_{l}$ are the temperatures of the two different heat
baths, and%
\begin{equation}
S(i)=-k_{B}\sum_{n}\frac{\exp [-\beta _{i}E_{n}(i)]}{Z(i)}[-\beta
_{i}E_{n}(i)-\ln Z(i)],  \label{9.9}
\end{equation}%
are the entropies of the working substance at different instants $i=A$, $B$, 
$C$, $D$ (see Fig.~3(A)). Here, $\beta _{A,B}=1/k_{B}T_{h}$, $\beta
_{C,D}=1/k_{B}T_{l}$. In obtaining the above result, we have used the
Boltzmann distribution of thermal equilibrium state, i.e., $\rho
=(1/Z)\sum_{n}\exp (-\beta E_{n})\left\vert n\right\rangle \left\langle
n\right\vert $, where $Z=$\textrm{$Tr$}$\exp (-\beta H)$ is the partition
function. Of course, $Q_{\mathrm{in}}^{\mathrm{QIT}}$ and $Q_{\mathrm{out}}^{%
\mathrm{QIT}}\,$\ can also be obtained through Eq.~(\ref{3}) in a quantum
manner (this will become clear later on, see Eqs.~(\ref{24}) and (\ref{25}%
)). These equivalent approaches can describe the microscopic mechanism of a
classical Carnot engine cycle.

Now, we would like to calculate the work $W_{\mathrm{C}}$ done during a QCE
cycle and its operation efficiency $\eta _{\mathrm{C}}$. From Eqs.~(\ref{9.3}%
) and (\ref{9.6}) and the first law of thermodynamics we obtain the net work
done during a QCE cycle%
\begin{equation}
W_{\mathrm{C}}=Q_{\mathrm{in}}^{\mathrm{QIT}}-Q_{\mathrm{out}}^{\mathrm{QIT}%
}=(T_{h}-T_{l})[S(B)-S(A)],  \label{9.8}
\end{equation}%
where we have used the relations $S(B)=S(C)$ and $S(A)=S(D)$. This
equivalence is due to the fact that the occupation probabilities and thus
the entropy remain invariant in any quantum adiabatic process. The
efficiency $\eta _{\mathrm{C}}$ of the QCE is%
\begin{equation}
\eta _{\mathrm{C}}\ =\ \frac{W_{\mathrm{C}}}{Q_{\mathrm{in}}^{\mathrm{QIT}}}%
\ =\ 1-\frac{T_{l}}{T_{h}},  \label{10}
\end{equation}%
which is just the efficiency of a classical Carnot engine. From Eq.~(\ref%
{10.1}) we see that the ratio of the temperature in the efficiency (Eq.~(\ref%
{10})) of the QCE can also be replaced by the ratio of the energy gaps 
\begin{equation}
\eta _{\mathrm{C}}\ =\ 1-\frac{E_{n}(C)-E_{m}(C)}{E_{n}(B)-E_{m}(B)}.
\label{10.2}
\end{equation}%
This expression of the efficiency $\eta _{\mathrm{C}}\ $in terms of the
ratio of the energy gaps resembles that of a QOE for a multi-level case in
Refs. \cite{kieu1,quan1} (see also Eq.~(\ref{13}) below). However, in spite
of the apparent similarities between these two expressions for the
efficiencies, we emphasize that they are quite different. Here, $%
E_{n}(B)-E_{m}(B)$ and $E_{n}(C)-E_{m}(C)$, in Eq.~(\ref{10.2}), are the
energy gaps at the beginning ($B$) and at the end ($C$) of the quantum
adiabatic expansion process ($B\longrightarrow C$). In the expression for
the efficiency $\eta _{\mathrm{O}}$ for a multi-level QOE, however, the
energy gaps are those in two quantum isochoric processes \cite{kieu1,quan1}.
Hence, the efficiency in Eq.~(\ref{10.2}) for a QCE is quite different from
that for a QOE, even though they both look similar. Further discussions on
this will be given in Sec. IV.

In order to extract positive work from the bath, Eq.~(\ref{10}) imposes a
constraint, $T_{h}>T_{l}$, on the temperatures of the two heat baths. This
constraint, known as the positive-work condition (PWC), is the same as that
of its classical counterpart. What is more, the schematic
temperature-entropy ($T-S$) diagrams for both a QCE cycle and a classical
Carnot engine cycle are the same (see Fig. 2(C)). For the above reasons, we
believe it is convincible that our QCE model is a quantum mechanical
analogue of a classical Carnot engine. We compare the properties of a QCE
and a classical Carnot engine and list them in Table II.

%\end{widetext}

\subsection{Internal energy}

It is well known that an ideal classical Carnot engine cycle consists of two
classical isothermal and two classical adiabatic processes. When the working
substance is the ideal gas, the internal energy of the working substance
remains invariant in the classical isothermal process, because the internal
energy of the ideal gas depends on the temperature only. This assumption for
classical isothermal processes based on classical ideal gas could be true
for a classical Carnot engine using a working substance other than an ideal
gas. But in the quantum version, the quantum isothermal and quantum
adiabatic processes should be redefined microscopically based on quantum
mechanics. In principle, the classical result could change when considering
the quantum nature (discrete energy levels) of the working substance.

We now would like to verify whether the internal energy of the working
substance remains invariant during the isothermal process. At the four
instants $A,B,C$ and $D$ of the QCE cycle (see Fig. 2), the internal
energies are respectively%
\begin{equation}
U(i)\ =\ \mathrm{Tr}[\rho (i)H(i)],\ \ \ \text{ }i\ =\ A,B,C,D.  \label{7}
\end{equation}%
In appendix B we prove that $U(A)\neq U(B)$ and $U(C)\neq U(D)$ for some QCE
cycles based on several experimentally realizable systems. Hence, in the
quantum version of a Carnot engine, we cannot simply assume that the heat
absorbed (released) by the working substance equals to the work done by (on)
the working substance in the isothermal process, as we do in classical
Carnot engine with the ideal gas as the working substance. This observation
is crucial for the following discussion.

Here we would like to indicate that, the quantum isoenergetic process in
Refs.~\cite{bender1,bender2} is not a quantum analogue of the classical
isothermal process of classical Carnot engine, because it requires the
temperature of the heat bath to change. Thus the thermodynamic cycle
described in Refs. \cite{bender1,bender2} is actually not a QCE cycle.

\section{QUANTUM OTTO ENGINE CYCLE}

In practice, the heat engines most widely used in automobiles, the internal
combustion engine, operate using Otto-cycle engines \cite{quan4}, which
consist of two classical isochoric and two classical adiabatic processes.
Similar to the Carnot engine, the quantum analogue of the classical Otto
engine is also proposed in Refs. \cite%
{kieu1,scully-pra,opatrny,quan1,scully2}. The QOE cycle consists of two
quantum isochoric and two quantum adiabatic processes \cite%
{kieu2,kosloff1,kosloff2,quan1} (see Fig. 3 for a schematic diagram of QOE
based on a two-level system).

%Figure 3
%\begin{figure}[h]
\begin{figure}[tbp]
\includegraphics[bb=178 195 409 726, clip, width=7cm]{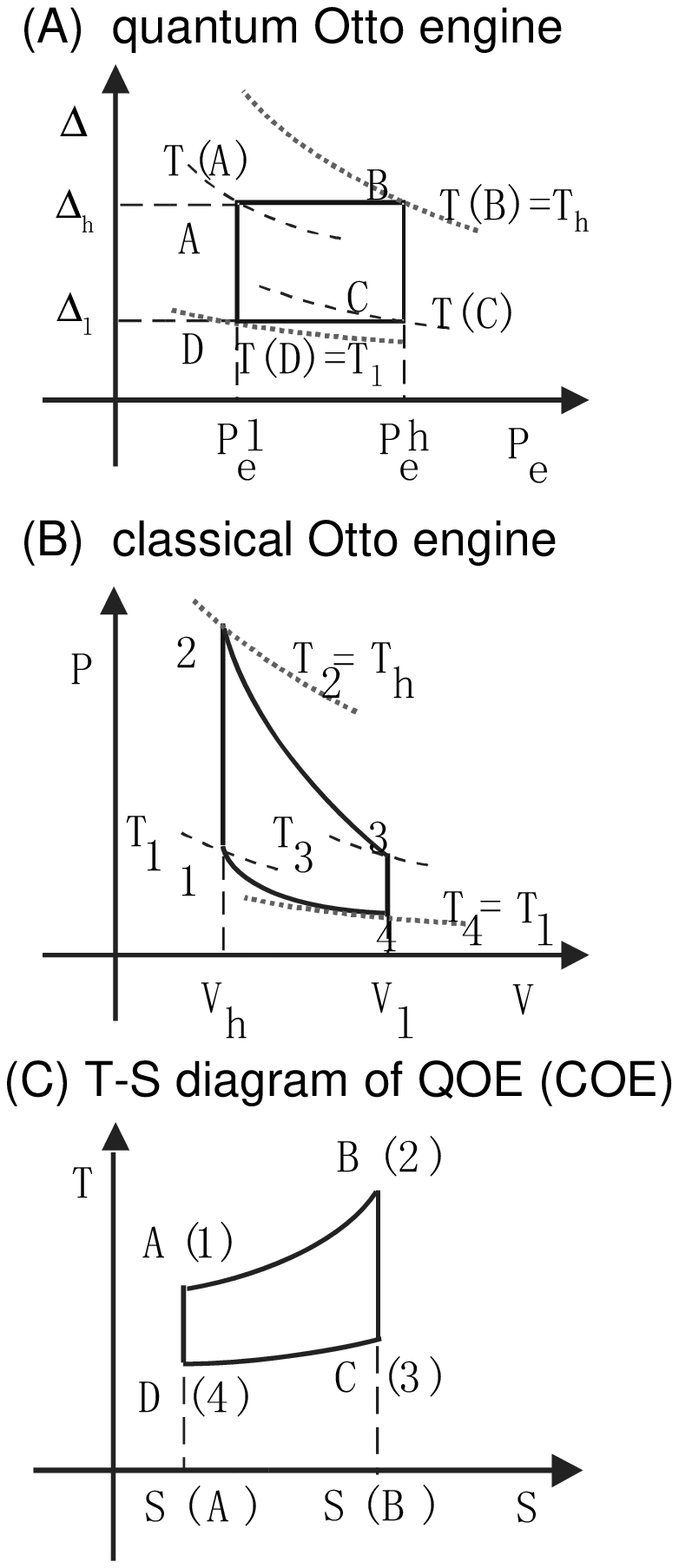}
\caption{(Color online) (A): A schematic diagram of a \textit{quantum} Otto
engine based on a two-level quantum system. Dotted red and dashed black
curves refer to the isothermal processes. $\Delta _{h}$ and $\Delta _{l}$
are the level spacings of the two-level system during the two isochoric
processes ($A\longrightarrow B$ and $C\longrightarrow D$). $P_{e}^{h}$ and $%
P_{e}^{l}$ are the occupation probabilities in the excited state. The
processes from $A$ to $B$ and from $C$ to $D$ are isochoric processes, while
the processes from $B$ to $C$ and from $D$ to $A$ are quantum adiabatic
process. $T(i)$, ($i=A$, $B$, $C$, $D$) are the effective temperatures of
the working substance at instant $i$, and $T(B)=T_{h}$, $T(D)=T_{l}$. \ (B):
Pressure-Volume, $PV$, diagram of a \textit{classical} Otto engine. $V_{h}$
and $V_{l} $ are the volume of the working substance in the two classical
isochoric processes. The process from $1$ to $2$ ($3$ to $4$) is a classical
isochoric heating (cooling) process, and the process from $2$ to $3 $ ($4$
to $1$) is the classical adiabatic expansion (compression) process. The
temperature of the working substance at $2$ and $4$ equal to the
temperatures of the two heat baths, $T_{h}$ and $T_{l}$, respectively. (C):
Temperature-Entropy ($T-S$) diagram \protect\cite{schroeder} for both for a 
\textit{quantum} Otto engine based on a two-level quantum system and a 
\textit{classical} Otto engine with ideal gas as the working substance. This 
$T-S$ diagram bridges the \textit{quantum} and \textit{classical} Otto
engine. }
\end{figure}

\subsection{Work and efficiency}

In the quantum isochoric heating process from $A$ to $B$ (see Fig. 3), no
work is done, but heat is absorbed. The heat $Q_{\mathrm{in}}^{\mathrm{QIC}}$
absorbed by the working substance is%
\begin{equation}
Q_{\mathrm{in}}^{\mathrm{QIC}}=\sum_{n}\int_{A}^{B}E_{n}\,dP_{n}=%
\sum_{n}E_{n}^{h}\,[P_{n}(B)-P_{n}(A)],  \label{11}
\end{equation}%
where $E_{n}^{h}$ is the $n$-th eigen energy of the system in the quantum
isochoric heating process from $A$ to $B$. Similarly, we obtain the heat
released to the low temperature entropy sink in the quantum isochoric
cooling process from $C$ to $D$ 
\begin{equation}
Q_{\mathrm{out}}^{\mathrm{QIC}}=-\sum_{n}\int_{C}^{D}E_{n}\,dP_{n}=%
\sum_{n}E_{n}^{l}\,[P_{n}(C)-P_{n}(D)],  \label{12}
\end{equation}%
where $E_{n}^{l}$ is the $n$-th eigen energy of the system in the quantum
isochoric cooling process. We would like to point out that in calculating $%
Q_{\mathrm{in}}^{\mathrm{QIC}}$ and $Q_{\mathrm{out}}^{\mathrm{QIC}}$, we
cannot apply Eqs.~(\ref{9.3}) and (\ref{9.6}), because $\dbarit Q=TdS$ is
only applicable to thermal equilibrium case, while in the quantum isochoric
process, the heat bath and the working substance are not always in thermal
equilibrium, i.e., this process is not thermodynamically reversible (for a
detailed discussion see below).

As mentioned above, in order to construct a QCE, all energy gaps must be
changed by the same ratio in quantum adiabatic process. But for a
multi-level QOE, there is no such a constraint (see Ref. \cite{quan1})
because we do not have to ensure the reversibility of the QOE cycle.
Nevertheless, to compare the QOE with the QCE, we only consider a special
case of QOE where all its energy gaps change by the same ratios as in the
quantum adiabatic processes, i.e., $E_{n}^{h}-E_{m}^{h}=\alpha
(E_{n}^{l}-E_{m}^{l})$, $(n=0,1,2,\ldots )$. When we choose $%
E_{0}^{h}=E_{0}^{l}=0$, i.e., the ground state eigen energies as the energy
reference point, we have $E_{n}^{h}=\alpha E_{n}^{l}$. Similar to the QCE,
the occupation distribution remains invariant in the two quantum adiabatic
processes, i.e., $P_{n}(B)=P_{n}(C)$ and $P_{n}(A)=P_{n}(D)$, and
accordingly the entropy remains invariant in the quantum adiabatic processes 
$S(B)=S(C)$ and $S(A)=S(D)$.

Based on this fact, and Eqs.~(\ref{11}) and (\ref{12}), we obtain the net
work $W_{\mathrm{O}}$ done during a QOE cycle%
\begin{equation}
W_{\mathrm{O}}=Q_{\mathrm{in}}^{\mathrm{QIC}}-Q_{\mathrm{out}}^{\mathrm{QIC}%
}=\sum_{n}(E_{n}^{h}-E_{n}^{l})\,[P_{n}(B)-P_{n}(A)].  \label{13.5}
\end{equation}%
and the operation efficiency $\eta _{\mathrm{O}}$ of the QOE cycle%
\begin{equation}
\eta _{\mathrm{O}}\ =\ \frac{W_{\mathrm{O}}}{Q_{\mathrm{in}}^{\mathrm{QIC}}}%
\ =1-\frac{E_{n}^{l}-E_{m}^{l}}{E_{n}^{h}-E_{m}^{h}}=\ 1-\frac{1}{\alpha }.
\label{13}
\end{equation}%
Here, $\alpha >1$ since $E_{n}^{h}>E_{n}^{l}$. This result, which stands for
a special multi-level QOE (all energy gaps change by the same ratio in the
quantum adiabatic process), is a generalization of the two-level QOE \cite%
{kieu1,quan1}. Let us recall the PWC of the special multi-level QOE \cite%
{quan1} mentioned above. From Eq.~(\ref{13.5}), the PWC for the special
multi-level QOE is 
\begin{equation}
T_{h}\ >\ \alpha \ T_{l}.  \label{14}
\end{equation}%
This is obviously different from that of a QCE, where the PWC is simply $%
T_{h}>T_{l}$.

Note that the first QHE model, initially proposed in Ref. \cite{scovil1}, is
actually a QOE, because its efficiency and its PWC are given by $\eta =1-\nu
_{1}/\nu _{p}$ and $T_{1}>(\nu _{p}/\nu _{1})T_{0}$, where $\nu _{1}$ and $%
\nu _{p}$ are the two energy gaps of the working substance, and $T_{1}$ and $%
T_{0}$ are the temperatures of the two heat baths, respectively. 
\begin{table*}[tbp]
\caption{Quantum Otto engine versus classical Otto engine. Here
\textquotedblleft CIC" and \textquotedblleft CA" refer to \textquotedblleft
Classical Iso-Choric processes" and \textquotedblleft Classical Adiabatic
processes", respectively; \textquotedblleft QIC" and \textquotedblleft QA"
refer to \textquotedblleft Quantum Iso-Choric processes" and
\textquotedblleft Quantum Adiabatic processes", respectively. Also, $V_{h}$
and $V_{l}$ are the volumes of the working substance (classical ideal gas)
in the two classical isochoric processes; $\protect\gamma $ is the classical
adiabatic exponent \protect\cite{schroeder}. $T(i)$ ($i=A$, $B$, $C$, $D$)
and $T_{k}$ ($k=1$, $2$, $3$, $4$) are defined in Fig.~3}
\begin{center}
\begin{tabular}{c|c|c|c}
\hline\hline
\parbox{2cm} {\ } & strokes & efficiency & 
\parbox{4cm} {positive-work
condition} \\ \hline\hline
Classical & 
\parbox{3cm} {\ \ \ \ \ \ \ \ \ \ \ \ \ \ \ \ \ \ \ \ \ \ \ \ \ \ \ \ \ \ \ \ \ \ \ \ \ \ \ \ CIC-CA-CIC-CA\
\ \ \ \ \ \ \ \ \ \ \ \ \ \ \ \ \ \ \ \ \ \ \ \ \ \ \ \ \ \ \ \ \ \
\ \ \ \ \ } & 
\parbox{7cm} {$\eta
_{\mathrm{O}}^{\mathrm{CL}}\ =\ 1-(\frac{V_{h}}{V_{l}})^{\gamma
-1}$$=1-\frac{T(C)}{T(B)}=1-\frac{T(D)}{T(A)}$} & $T_{h}>T_{l}(\frac{V_{l}}{%
V_{h}})^{\gamma -1}$ \\ \hline
Quantum & 
\parbox{3cm}
{\ \ \ \ \ \ \ \ \ \ \ \ \ \ \ \ \ \ \ \ \ \ \ \ \ \ \ \ \ \ \ \ \ \
\ \ \ \ \ \ QIC-QA-QIC-QA\ \
 \ \ \ \ \ \ \ \ \ \ \ \ \ \ \ \ \ \ \ \ \ \ \ \ \ \ \ \ \ \ \ \ \ \ \ \ \
  \ } & 
\parbox{7cm} {$\eta _{\mathrm{O}}=1-\frac{\Delta _{l}}{\Delta _{h}}$
$=1-\frac{T_{3}}{T_{2}}=1-\frac{T_{4}}{T_{1}}$} & $T_{h}>T_{l}(\frac{\Delta
_{h}}{\Delta _{l}}) $ \\ \hline\hline
\end{tabular}%
\end{center}
\end{table*}

\subsection{Classical versus quantum Otto engines}

Below we prove that the operation efficiency $\eta _{\mathrm{O}}$ in Eq.~(%
\ref{13}) of a QOE, also equals to the efficiency $\eta _{\mathrm{O}}^{%
\mathrm{CL}}$ of a classical Otto engine. For simplicity, here we only
consider a two-level system as the working substance (the result can be
generalized to multi-level systems if all the eigen energies of the
multi-level system change in the same ratios as in the quantum adiabatic
process \cite{explain}). For a two-level system (see Fig. 3), when the
temperature $T$ of the heat bath is fixed, the occupation probability $P_{e}$
of the excited state $\left\vert e\right\rangle $ in thermal equilibrium is
a monotonically decreasing function of the level spacing $\Delta $ between
the two energy levels \cite{kieu1,arnaud1,arnaud2}. Its inverse function
reads%
\begin{equation}
\Delta _{\theta }=k_{B}T_{\theta }\ln \left[ \frac{1}{P_{e}^{\theta }}-1%
\right] ,\ \ \ \ \theta =h,l.  \label{16}
\end{equation}%
As mentioned above, the efficiency $\eta _{\mathrm{O}}$ (\ref{13}) of a QOE
cycle represented by the rectangle ($A$-$B$-$C$-$D$) (see Fig. 3) is%
\begin{equation}
\eta _{\mathrm{O}}=1-\frac{\Delta _{l}}{\Delta _{h}}.  \label{17}
\end{equation}%
From Eqs.~(\ref{16}) and (\ref{17}) and Fig. 3, we can see that the
efficiency $\eta _{\mathrm{O}}$ of the QOE cycle can be rewritten as%
\begin{equation}
\eta _{\mathrm{O}}=1-\frac{T(C)}{T(B)}=1-\frac{T(D)}{T(A)},  \label{19}
\end{equation}%
where $T(i)$, with $i=A$, $B$, $C$ and $D$, are the effective temperatures
of the working substance at the instants $A$, $B$, $C$ and $D$ indicated in
Fig. 3. Here, we have used the relation $\Delta (C)/\Delta (B)=T(C)/T(B)$
because of the fact $P_{e}(C)=P_{e}(B)=P_{e}^{h}$ and Eq.~(\ref{16}), and
similarly $\Delta (D)/\Delta (A)=T(D)/T(A)$. In the QOE cycle, the effective
temperatures $T(B)$ and $T(D)$ of the working substance at instants $B$ and $%
D$ equal the temperatures of the two heat baths $T(B)=T_{h}$, $T(D)=T_{l}$.

As for a classical Otto engine, the classical efficiency $\eta _{\mathrm{O}%
}^{\mathrm{CL}}$ is \cite{schroeder} 
\begin{equation}
\eta _{\mathrm{O}}^{\mathrm{CL}}=1-\left( \frac{V_{h}}{V_{l}}\right)
^{\gamma -1},  \label{19.3}
\end{equation}%
where $V_{l}$ and $V_{h}$ are the volumes of the classical ideal gas in the
two classical isochoric processes (see Fig.~3), and $\gamma $ is the
classical adiabatic exponent \cite{schroeder}. Because $TV^{\gamma -1}$ is
constant during a classical adiabatic process, we can therefore eliminate
the volumes in Eq.~(\ref{19.3}) as 
\begin{equation}
\eta _{\mathrm{O}}^{\mathrm{CL}}=1-\frac{T_{3}}{T_{2}}=1-\frac{T_{4}}{T_{1}},
\label{19.6}
\end{equation}%
where $T_{1}$, $T_{2}$, $T_{3}$ and $T_{4}$ are the temperatures of the
working substance at instants $1$, $2$, $3$, and $4$, and the temperatures
at instants $2$ and $4$ equal to the temperatures of the two heat baths $%
T_{2}=T_{h}$, $T_{4}=T_{l}$. This is in very good agreement with the result
of a QOE cycle in Eq.~(\ref{19}). Thus we proved that the efficiency of a
QOE also equals its classical counterpart. Moreover, similar to Carnot
engines, we plot the schematic temperature-entropy ($T-S$) diagrams for both
a QOE cycle and a classical Otto engine cycle in Fig. 3(C). The similarity
of the two $T-S$ diagrams also support our definition of the QOE.
Comparisons between the QOE and the classical Otto engine are listed in
Table III. %\begin{widetext}

%\end{widetext}\ \

\subsection{An alternative quantum Otto engine}

Before concluding this section, we would like to revisit an alternative QOE
cycle similar to that in Ref. \cite{lloyd} and that given most recently by
us \cite{quan4}. As illustrated in Fig. 4, we consider two two-level systems
(two qubits) as the working substance, which are denoted by qubit $S$ and
qubit $D$, and the level spacings of the two qubits are $\Delta _{S}$ and $%
\Delta _{D}$, respectively. The temperatures of the two heat baths are $%
T_{S} $ and $T_{D}$. Without loss of generality, here we choose $T_{S}>T_{D}$
\ and $\Delta _{S}>\Delta _{D}$.

%Figure 9
%\begin{figure}[h]
\begin{figure}[tbp]
\begin{center}
\includegraphics[bb=44 290 572 697, width=8 cm, clip]{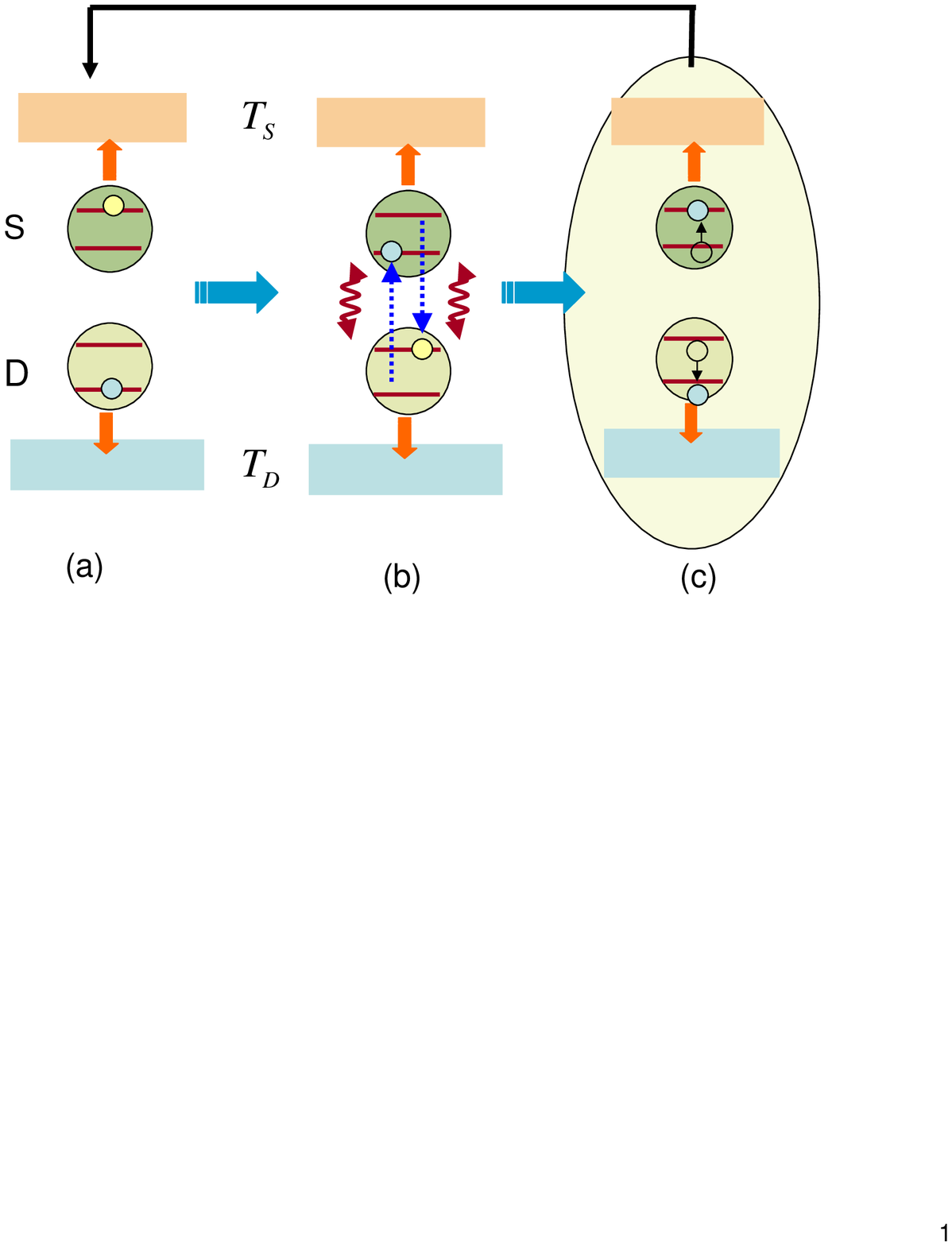}
\end{center}
\caption{(Color online) A schematic diagram illustrating an alternative
quantum Otto engine. This QOE cycle consists of two steps: a SWAP between
the two two-level systems (two qubits), shown in (b), and a thermalization
with their respective heat baths, shown in (a). Step (c) indicates the
transition from (b) to (a). The SWAP operation in (b) replaces the two
quantum isochoric processes in the QOE cycle mentioned in the text and shown
in Fig.~3(A).}
\end{figure}

The alternative QOE cycle consists of two steps: 1) let the two qubits
decouple from each other and contact their own heat baths respectively until
they reach thermal equilibrium with these two heat baths; and 2) switch on
the interaction between the two qubits and implement a SWAP operation
between them. These two steps are shown in Fig.~4(a) and 4(b). The density
matrices of the two qubits after step 1 are%
\begin{equation}
\rho _{i}(1)=\frac{1}{Z_{i}}\left[ \left\vert 0\right\rangle
_{i}\left\langle 0\right\vert _{i}+\exp [-\beta _{i}\Delta _{i}]\left\vert
1\right\rangle _{i}\left\langle 1\right\vert _{i}\right] ,\ \ (i=S,D),
\label{19.7}
\end{equation}%
where $\beta _{i}=1/k_{B}T_{i}$, and for simplicity we have chosen the eigen
energy of the ground state as a reference point. After step 2, the density
matrices become%
\begin{eqnarray}
\rho _{S}(2) &=&\frac{1}{Z_{D}}\left[ \left\vert 0\right\rangle
_{S}\left\langle 0\right\vert _{S}+\exp [-\beta _{D}\Delta _{D}]\left\vert
1\right\rangle _{S}\left\langle 1\right\vert _{S}\right] ,  \notag \\
\rho _{D}(2) &=&\frac{1}{Z_{S}}\left[ \left\vert 0\right\rangle
_{D}\left\langle 0\right\vert _{D}+\exp [-\beta _{S}\Delta _{S}]\left\vert
1\right\rangle _{D}\left\langle 1\right\vert _{D}\right] .  \label{19.8}
\end{eqnarray}%
After these two steps, the two qubits are decoupled and put into contact
with their own bath, and a new cycle starts.

The key point of this alternative QOE cycle is that the SWAP operation takes
place of the two quantum adiabatic processes, while the thermolization
process takes place of two quantum isochoric processes. We can now calculate
the heat absorbed by the qubit $S$ and heat released by the qubit $D$ in
step 1:%
\begin{eqnarray}
Q_{\mathrm{in}} &=&\mathrm{Tr}[H\rho _{S}(1)]-\mathrm{Tr}[H\rho _{S}(2)]
\label{19.4} \\
&=&\Delta _{S}\left[ \frac{1}{Z_{S}}\exp [-\beta _{S}\Delta _{S}]-\frac{1}{%
Z_{D}}\exp [-\beta _{D}\Delta _{D}]\right] ,  \notag
\end{eqnarray}%
\begin{eqnarray}
Q_{\mathrm{out}} &=&\mathrm{Tr}[H\rho _{D}(2)]-\mathrm{Tr}[H\rho _{D}(1)]
\label{19.5} \\
&=&\Delta _{D}\left[ \frac{1}{Z_{S}}\exp [-\beta _{S}\Delta _{S}]-\frac{1}{%
Z_{D}}\exp [-\beta _{D}\Delta _{D}]\right] .  \notag
\end{eqnarray}%
The operation efficiency $\eta $ of the QHE cycle can be calculated
straightforwardly%
\begin{equation}
\eta =\frac{Q_{\mathrm{in}}-Q_{\mathrm{out}}}{Q_{\mathrm{in}}}=1-\frac{%
\Delta _{D}}{\Delta _{S}}  \label{19.1}
\end{equation}%
and the PWC is $T_{S}>(\Delta _{S}/\Delta _{D})T_{D}$. Thus, this two-step
cycle is an alternative QOE cycle. We will revisit, in more detail, this
alternative QOE cycle in Section VII.

Let us here mention an alternative QCE of two qubits based on a SWAP
operation. This alternative QCE cycle consists of three steps: 1) let the
two qubits decouple and contact their own heat baths and both experience
quantum isothermal processes, 2) switch on the interaction between the two
qubits and implement a SWAP operation, and 3) let the two qubits decouple
from each other and also decouple the two qubits from their heat baths, and
subject them to a quantum adiabatic process. A similar QOE cycle and a QCE
cycle have been studied in Ref. \cite{lloyd}, where the SWAP operation was
decomposed into three CNOT operations.

\section{RELATIONS BETWEEN QUANTUM OTTO AND QUANTUM CARNOT CYCLES}

In this section we discuss the relation between a quantum Otto engine cycle
and a quantum Carnot engine cycle. For simplicity, here we use a two-level
system as an example of working substance. Our results about QHEs based on a
two-level system can be generalized to multi-level systems if all the eigen
energies of the multi-level system change by the same ratios in the quantum
adiabatic processes \cite{explain}.

\subsection{Quantum Carnot cycle derived from quantum Otto cycles}

A QCE cycle can be decomposed into an infinite number of small QOE cycles
(see Fig. 5) \cite{kieu1,arnaud2,kosloff1,kosloff2}. Now we give a concise
and explicit proof about this observation. The heat absorbed and released in
the infinite number of infinitesimal QOE cycles can be integrated by
applying Eqs.~(\ref{11}), (\ref{12}) and (\ref{16})%
\begin{eqnarray}
Q_{\mathrm{in}} &=&T_{h}\int_{P_{e}^{l}}^{P_{e}^{h}}\,\ln \left(
P_{e}^{-1}-1\right) \ d\!P_{e},  \label{24} \\
Q_{\mathrm{out}} &=&T_{l}\int_{P_{e}^{l}}^{P_{e}^{h}}\,\ln \left(
P_{e}^{-1}-1\right) \ d\!P_{e}.  \label{25}
\end{eqnarray}%
Then we obtain the positive work $W$ done during the infinite number of
infinitesimal QOE cycles ($A$-$B$-$C$-$D$) by making use of Eq.~(\ref{13.5}) 
\begin{eqnarray}
W &=&Q_{\mathrm{in}}-Q_{\mathrm{out}}  \label{26} \\
&=&(T_{h}-T_{l})\int_{P_{e}^{l}}^{P_{e}^{h}}\,\ln \left( P_{e}^{-1}-1\right)
\ d\!P_{e}.  \notag
\end{eqnarray}%
From Eqs.~(\ref{24}) and (\ref{26}) we can see that the efficiency of an
infinite number of infinitesimal QOE cycles is%
\begin{equation}
\eta =\frac{W}{Q_{\mathrm{in}}}=1-\frac{T_{l}}{T_{h}}  \label{26.1}
\end{equation}%
This is the efficiency of the QCE in Eq.~(\ref{10}). Thus we have proved
that a QCE cycle can be modeled as an infinite number of infinitesimal QOE
cycles. However, a finite QCE cycle and a finite QOE cycle cannot be
equivalent because one is reversible and the other one is not. When the two
cycles becomes infinitesimal, they can be infinitesimally close to each
other. %Figure 5
%\begin{figure}[h]
\begin{figure}[tbp]
\begin{center}
\includegraphics[bb=46 361 556 751, width=8cm, clip]{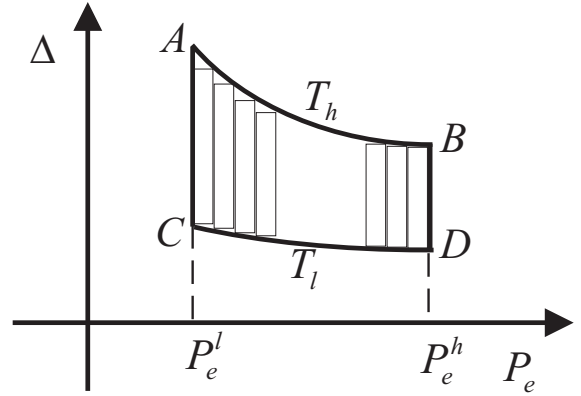}
\end{center}
\caption{A Quantum Carnot engine cycle can be modeled as infinite number of
small quantum Otto engine. Here, $P_{e}^{h}$, $P_{e}^{l}$,$A$, $B$, $C$, $D$%
, and $T_{h}$, $T_{l}$ are defined in Fig.~2(A). The small rectangles inside
the QCE cycle ($A$-$B$-$C$-$D$) represent small QOE cycles. The temperatures
of the two heat baths of these QOE cycles are $T_{h}$ and $T_{l}$,
respectively, which are the same as that of the QCE cycle. Similar
discussions see also Refs.~\protect\cite{kieu1,arnaud2,kosloff1,kosloff2}}
\end{figure}

\subsection{Quantum Otto cycle derived from quantum Carnot cycles}

Conversely, a QOE cycle can also be modelled as an infinite number of QCE
cycles (see Fig. 6), but the temperatures of the two heat baths of these
small QCE are different. The quantum isochoric process $A^{\prime
}\longrightarrow B^{\prime }$ ($C^{\prime }\longrightarrow D^{\prime }$) can
be modelled as many small quantum isothermal processes with temperatures $%
T_{h}^{1}$ ($T_{l}$), $T_{h}^{2}$ ($T_{l}^{1}$), $T_{h}^{3}$ ($T_{l}^{2}$), $%
\ldots $(see Fig. 6) 
\begin{eqnarray}
T_{h}^{1} &<&T_{h}^{2}<T_{h}^{3}<\cdots <T_{h}^{N}<T_{h},  \label{28} \\
T_{l} &<&T_{l}^{1}<T_{l}^{2}<\cdots <T_{l}^{N-1}<T_{l}^{N}.  \label{29}
\end{eqnarray}%
The heat absorbed and released in the infinite number of infinitesimal QCE
cycles can be obtained by applying $\dbarit Q=TdS$ and Eq. (\ref{16})%
\begin{equation}
Q_{\mathrm{in}}=\int_{P_{e}^{l}}^{P_{e}^{h}}\frac{\Delta _{h}}{k_{B}}\frac{1%
}{\ln \left( P_{e}^{-1}-1\right) }dS=\Delta
_{h}\int_{P_{e}^{l}}^{P_{e}^{h}}d\!P_{e},  \label{26.2}
\end{equation}%
\begin{equation}
Q_{\mathrm{out}}=\int_{P_{e}^{l}}^{P_{e}^{h}}\frac{\Delta _{l}}{k_{B}}\frac{1%
}{\ln \left( P_{e}^{-1}-1\right) }dS=\Delta
_{l}\int_{P_{e}^{l}}^{P_{e}^{h}}d\!P_{e},  \label{26.3}
\end{equation}%
where we have used the relation $dS=-k_{B}[\ln P_{e}-\ln (1-P_{e})]d\!P_{e}$%
. We then obtain the positive work $W$ done during the infinite number of
infinitesimal QCE cycles ($A^{\prime }$-$B^{\prime }$-$C^{\prime }$-$%
D^{\prime }$ in Fig. 6) by making use of Eq. (\ref{9.8})%
\begin{equation}
W=Q_{\mathrm{in}}-Q_{\mathrm{out}}=(\Delta _{h}-\Delta
_{l})(P_{e}^{h}-P_{e}^{l}).  \label{26.4}
\end{equation}%
From Eqs. (\ref{26.2}) and (\ref{26.4}) we can see that the efficiency of an
infinite number of infinitesimal QCE cycle is%
\begin{equation}
\eta =\frac{W}{Q_{\mathrm{in}}}=1-\frac{\Delta _{l}}{\Delta _{h}}.
\label{26.6}
\end{equation}%
This is the efficiency of the QOE in Eq. (\ref{17}). Thus we have proved
that a QOE cycle can be modeled as infinite number of infinitesimal QCE
cycles.

We would like to mention that the formula $\dbarit Q=TdS$ is applicable in
Eq. (\ref{26.2}) is due to the fact that these infinite number of
infinitesimal QCE cycles (with heat baths temperatures $T_{h}^{1}$, $%
T_{l}^{1}$; $T_{h}^{2}$, $T_{l}^{2}$; $\cdots $) are thermal equilibrium
(reversible) processes since the entropy increase vanishes during these
cycles (see Appendix C). However, if we take the QOE cycle as a whole, and
the heat baths temperatures are $T_{h}$ and $T_{l}$, this process is a
nonequilibrium (irreversible) process. Hence, the formula $\dbarit Q=TdS$ is
not applicable here. This is why we cannot apply Eqs.~(\ref{9.3}) and (\ref%
{9.6}) in calculating $Q_{\mathrm{in}}^{\mathrm{QIC}}$ and $Q_{\mathrm{out}%
}^{\mathrm{QIC}}$ in the QOE cycle in Sec. IVA.

%Figure 6
%\begin{figure}[h]
\begin{figure}[tbp]
\begin{center}
\includegraphics[bb=43 356 561 760, width=8cm, clip]{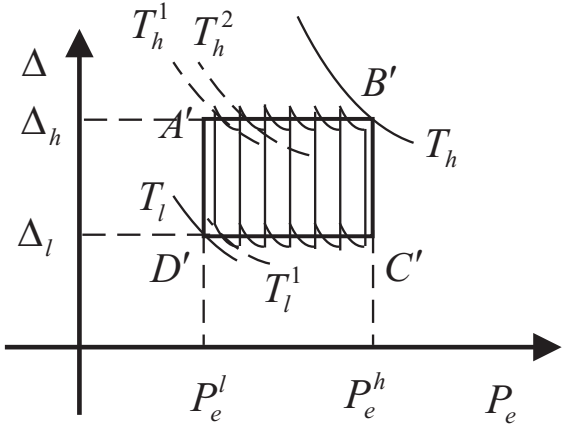}
\end{center}
\caption{A Quantum Otto engine can be modeled as an infinite number of small
quantum Carnot engine cycles. Here $P_{e}^{h}$, $P_{e}^{l}$, $T_{h}$ and $%
T_{l}$ are defined in Fig. 3(A). The small cycles inside the QOE cycle $%
A^{\prime }$-$B^{\prime }$-$C^{\prime }$-$D^{\prime }$ represent small
quantum Carnot cycles. The temperatures of the two heat baths (e.g., $%
T_{h}^{1}$ and $T_{l}^{1}$) of these small QCE cycles are different from
that ($T_{h}$ and $T_{l}$) of the QCE cycle.}
\end{figure}

\subsection{Comparison of work and efficiency for quantum Otto and quantum
Carnot cycles}

Having clarified the properties of the QCE and QOE, here we now compare the
thermodynamic properties of a QCE cycle with that of a QOE cycle and study
the relation between them. For the two QHE cycles (QCE cycle and QOE cycle),
we consider the case with the same heat baths (at high and low temperatures $%
T_{h}$ and $T_{l}$, respectively), and the same occupation probabilities ($%
P_{e}^{h}$ and $P_{e}^{l}$, respectively) in the two quantum adiabatic
processes (see Fig.~7).

First, let us compare the amounts of positive work done in a QCE cycle and a
QOE cycle under the same conditions defined above. The positive work done
during the QCE cycle (here denoted by $A$-$B^{\prime }$-$C$-$D^{\prime }$ in
Fig.~7) can also be written as%
\begin{equation}
W_{\mathrm{C}}=\int_{P_{e}^{l}}^{P_{e}^{h}}[\Delta (T_{h},P_{e})-\Delta
(T_{l},P_{e})]\ d\!P_{e},  \label{26.5}
\end{equation}%
while the work done during a QOE cycle is (here denoted by $A^{\prime }$-$%
B^{\prime }$-$C^{\prime }$-$D^{\prime }$ in Fig.~7)%
\begin{equation}
W_{\mathrm{O}}=\int_{P_{e}^{l}}^{P_{e}^{h}}(\Delta _{h}-\Delta _{l})\
d\!P_{e}.  \label{27}
\end{equation}%
From Fig.~7 we know for any $P_{e}\in \lbrack P_{e}^{l},P_{e}^{h}]$, we have 
$\Delta (T_{h},P_{e})-\Delta (T_{l},P_{e})>\Delta _{h}-\Delta _{l}$. Hence,
from Eqs.~(\ref{26.5}) and (\ref{27}) we have $W_{\mathrm{C}}>W_{\mathrm{O}}$%
; i.e., under the same conditions, the work done during a QCE cycle is more
than that during a QOE cycle.

%Figure 4
%\begin{figure}[h]
\begin{figure}[tbp]
\begin{center}
\includegraphics[bb=61 408 541 756, width=8cm, clip]{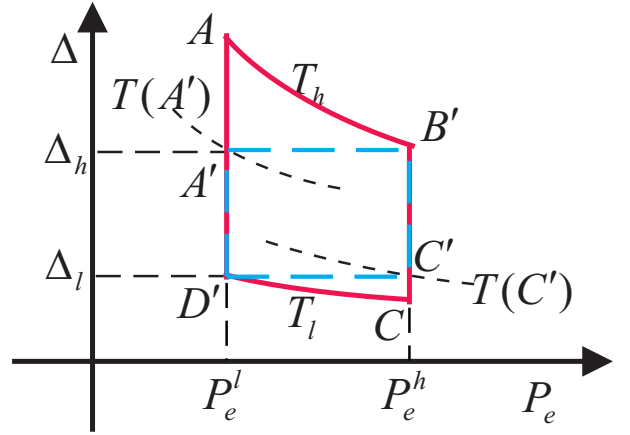}
\end{center}
\caption{(Color online) Schematic diagrams for a quantum Carnot engine cycle
($AB^{\prime }CD^{\prime }$ red continuous line) and a quantum Otto engine
cycle ($A^{\prime }B^{\prime }C^{\prime }D^{\prime }$ blue dashed line)
based on a two-level system. Here, points $B^{\prime }$ and $D^{\prime }$
are the same for both cycles. The temperatures $T_{h}$ and $T_{l}$ of the
two heat baths of the two QHE cycles and the occupation probabilities $%
P_{e}^{h}$ and $P_{e}^{l}$ in two quantum adiabatic processes are the same.
The net work done during a cycle is proportional to the area enclosed by the
four curves which represent the cycles.}
\end{figure}

Next we consider the efficiencies of the QCE cycle and the QOE cycle under
the same conditions. From Eq.~(\ref{17}) we know that the efficiency of a
small QOE cycle can be rewritten as%
\begin{eqnarray}
\eta _{\mathrm{O}} &=&1-\frac{k_{B}T_{l}\ln [(P_{e}^{l})^{-1}-1]}{%
k_{B}T_{h}\ln [(P_{e}^{l})^{-1}-1]}  \label{27.1} \\
&\simeq &1-\frac{T_{l}}{T_{h}}\left[ 1-\frac{\partial }{\partial P_{e}}\left[
\ln \ln \left( \frac{1}{P_{e}^{l}}-1\right) \right] (P_{e}^{h}-P_{e}^{l})%
\right]  \notag \\
&=&\eta _{\mathrm{C}}+\frac{T_{l}}{T_{h}}\frac{\partial }{\partial P_{e}}%
\left[ \ln \ln \left( \frac{1}{P_{e}^{l}}-1\right) \right]
(P_{e}^{h}-P_{e}^{l}).  \notag
\end{eqnarray}%
It can be verified that the second term on the RHS of Eq. (\ref{27.1}) is
negative. Thus we have proved the inequality%
\begin{equation}
\eta _{\mathrm{O}}\ =\ 1-\frac{\Delta _{l}}{\Delta _{h}}\ <\ 1-\frac{T_{l}}{%
T_{h}}\ =\ \eta _{\mathrm{C}}  \label{15}
\end{equation}%
for every small cycle. We conclude that, under the same conditions, the QCE
is more efficient that the QOE, even for any finite cycle.

\section{ILLUSTRATIONS OF QUANTUM CARNOT ENGINE AND QUANTUM OTTO ENGINE}

As mentioned above, to construct a multi-level (including two-level)
Carnot-type QHE, two preconditions (mathematical results) are required: (1)
all energy gaps change by the same ratio in the quantum adiabatic process
(when the working substance performs work); (2) the ratio of the energy gap
changes in the quantum adiabatic process should equal the ratio of the
temperatures of the two heat baths, so that the thermodynamic cycle is
reversible. Physically, these two conditions can always be satisfied for a
QHE based on a two-level system, because there is only one energy gap in a
two-level system, we can always find a proper effective temperature to
characterize the working substance. Besides the two-level system, the
harmonic oscillator and a\ particle confined in an infinite square well
potential are two other examples that can illustrate the basic properties of
the QCE and the QOE. This is because in both cases all energy gaps change by
the same ratio when changing the parameters of the system, and we can always
use a proper effective temperature to characterize the working substance in
the quantum adiabatic process, too. Below we calculate the amount of
positive work done during a thermodynamic cycle using those working
substances. 
\begin{table*}[tbp]
\caption{Comparison of several properties of quantum Otto engines and
quantum Carnot engines using the following \textit{quantum} substances as
working substances: Two-Level System (TLS), Harmonic Oscillator (HO), and a
particle confined in an Infinite Square (IS) potential. The operation
efficiency $\protect\eta $, the positive-working condition (PWC) and the
amount of work $W$ extracted in a thermodynamic cycle are listed below (see
Sec. VI).}
\begin{center}
\begin{tabular}{|c|c|c|c|c|}
\hline\hline
\multicolumn{2}{|c|}{} & \parbox{4cm} {Two-Level System
(TLS)} & \parbox{4.5cm} {Harmonic Oscillator (HO)} & 
\parbox{4cm}
{Infinite Square well (IS)} \\ \hline\hline
& \parbox{1cm}
{$\eta_{\mathrm{O}}$} & $1-\frac{\Delta _{l}}{\Delta _{h}}$ & $1-\frac{%
\omega _{l}}{\omega _{h}}$ & $1-\left( \frac{L_{l}}{L_{h}}\right) ^{2}$ \\ 
\cline{2-5}
\parbox{2cm} {Quantum Otto engine} & PWC & $T_{h}>\frac{\Delta _{h}}{\Delta
_{l}}T_{l}$ & $T_{h}>\frac{\omega _{h}}{\omega _{l}}T_{l}$ & $T_{h}>\left( 
\frac{L_{h}}{L_{l}}\right) ^{2}T_{l}$ \\ \cline{2-5}
& $W_{\mathrm{O}}$ & $W_{\mathrm{O}}^{\mathrm{TLS}}$ & $W_{\mathrm{O}}^{%
\mathrm{HO}}$ & $W_{\mathrm{O}}^{\mathrm{IS}} $ \\ \hline
& $\eta_{\mathrm{C}}$ & $1-\frac{T_{l}}{T_{h}}$ & $1-\frac{T_{l}}{T_{h}}$ & $%
1-\frac{T_{l}}{T_{h}}$ \\ \cline{2-5}
\parbox{2cm} {Quantum Carnot engine} & PWC & $T_{h}>T_{l}$ & $T_{h}>T_{l}$ & 
$T_{h}>T_{l}$ \\ \cline{2-5}
& $W_{\mathrm{C}}$ & $W_{\mathrm{C}}^{\mathrm{TLS}}$ & $W_{\mathrm{C}}^{%
\mathrm{HO}}$ & $W_{\mathrm{C}}^{\mathrm{IS}}$ \\ \hline\hline
\end{tabular}%
\end{center}
\end{table*}

\subsection{Two-level systems}

Let us now consider a QHE based on a two-level system, e.g., a spin-$1/2$
system in an external magnetic field pointing along the $+z$ direction, the
Hamiltonian of the working substance is%
\begin{equation}
H_{\mathrm{TLS}}(i)=-MB(i)(\left\vert \uparrow \right\rangle \left\langle
\uparrow \right\vert -\left\vert \downarrow \right\rangle \left\langle
\downarrow \right\vert ),  \label{29.1}
\end{equation}%
where $B(i)$ is the strength of the external field at instant $i$, $%
i=A,B,C,D $ (see Fig. 2 and Fig. 3); $M=e\hbar /2mc$ is the Bohr magnon; $%
\left\vert \downarrow \right\rangle $ and $\left\vert \uparrow \right\rangle 
$ indicate the spin down (excited) state and spin up (ground) state
respectively. The thermal equilibrium state can be written as%
\begin{eqnarray}
\rho _{\mathrm{TLS}}(i) &=&\frac{1}{Z(i)}[\exp [\beta _{i}MB(i)]\left\vert
\uparrow \right\rangle \left\langle \uparrow \right\vert  \label{29.2} \\
&&+\exp [-\beta _{i}MB(i)]\left\vert \downarrow \right\rangle \left\langle
\downarrow \right\vert ],  \notag
\end{eqnarray}%
where $Z(i)=\exp [\beta _{i}MB(i)]+\exp [-\beta _{i}MB(i)]$ is the partition
function at instant $i$. Applying Eq.~(\ref{9.9}) we obtain the entropy of
the working substance%
\begin{equation}
S^{\mathrm{TLS}}(i)=k_{B}\ln [Z(i)]+k_{B}\beta _{i}MB(i)\ \tanh [\beta
_{i}MB(i)].  \label{29.3}
\end{equation}%
Then from Eqs.~(\ref{9.3}), (\ref{9.6}) and (\ref{9.8}) we obtain the net
work done by the working substance during a QCE cycle%
\begin{equation}
W_{\mathrm{C}}^{\mathrm{TLS}}=(T_{h}-T_{l})\left[ S^{\mathrm{TLS}}(B)-S^{%
\mathrm{TLS}}(A)\right] .  \label{29.4}
\end{equation}%
The net work done during the QOE cycle can be calculated by applying Eq.~(%
\ref{13.5})%
\begin{equation}
W_{\mathrm{O}}^{\mathrm{TLS}}=(\Delta _{h}-\Delta _{l})\left[ \frac{1}{%
1+\exp [\beta _{h}\Delta _{h}]}-\frac{1}{1+\exp [\beta _{l}\Delta _{l}]}%
\right] .  \label{29.0}
\end{equation}%
\ \ \ 

Another example of QHE based on a two-level system is a photon-Carnot engine 
\cite{scully-science,quan2,pce}. After performing a similar calculation we
recover the operation efficiency $1-T_{l}/T_{h}$ in Eq.~(\ref{10}) and the
PWC $T_{h}>T_{l}$. Hence, this photon-Carnot engine is actually a two-level
QCE.

\subsection{Harmonic oscillator}

For a QHE based on a harmonic oscillator with the eigen energies $%
E_{n}(i)=(n-1/2)\hbar \omega (i)$, by applying Eq.~(\ref{9.9}), we derive
the entropy $S^{\mathrm{HO}}(i)$ of the working substance as%
\begin{eqnarray}
S^{\mathrm{HO}}(i) &=&-k_{B}\ln \{1-\exp [-\beta _{i}\hbar \omega (i)]\}
\label{29.5} \\
&&+k_{B}\beta _{i}\hbar \omega (i)\frac{1}{\exp [\beta _{i}\hbar \omega
(i)]-1},  \notag
\end{eqnarray}%
and the work done during a QCE cycle as%
\begin{equation}
W_{\mathrm{C}}^{\mathrm{HO}}=(T_{h}-T_{l})\left[ S^{\mathrm{HO}}(B)-S^{%
\mathrm{HO}}(A)\right] .  \label{29.6}
\end{equation}%
The net work done during the QOE cycle can be obtained by applying Eq. (\ref%
{13.5})%
\begin{equation}
W_{\mathrm{O}}^{\mathrm{HO}}=\hbar (\omega _{h}-\omega _{l})\left[ \frac{1}{%
\exp [\beta _{h}\hbar \omega _{h}]-1}-\frac{1}{\exp [\beta _{l}\hbar \omega
_{l}]-1}\right] .  \label{28.9}
\end{equation}

\subsection{Particle in an infinite square potential well}

For a QHE based on a particle confined in an Infinite Square (IS) potential,
the eigen energies are $E_{n}(i)=\gamma _{i}n^{2}$, where $\gamma _{i}=(\pi
\hbar )^{2}/(2mL_{i}^{2})$; $m$ and $L_{i}$ are the mass of the particle and
the width of the square well at instant $i$, respectively. The entropy of
the working substance can also be calculated by applying Eq.~(\ref{9.9}): 
\begin{equation}
S^{\mathrm{IS}}(i)=\frac{k_{B}}{2}(\beta _{i}\gamma _{i})^{\frac{3}{4}%
}+k_{B}\ln \left( \frac{1}{2}\sqrt{\frac{\pi }{\beta _{i}\gamma _{i}}}%
\right) .  \label{29.7}
\end{equation}%
In obtaining Eq.~(\ref{29.7}) we have make an approximation%
\begin{equation}
\sum_{n=1}^{\infty }\exp [\gamma _{i}n^{2}]\approx \int_{0}^{\infty }\exp
[\gamma _{i}n^{2}]dn.  \label{29.9}
\end{equation}%
So the work done during a QCE cycle can be expressed as (\ref{9.8})%
\begin{equation}
W_{\mathrm{C}}^{\mathrm{IS}}=(T_{h}-T_{l})\left[ S^{\mathrm{IS}}(B)-S^{%
\mathrm{IS}}(A)\right] ,  \label{29.8}
\end{equation}%
while the net work done during a QOE cycle is (\ref{13.5})%
\begin{equation}
W_{\mathrm{O}}^{\mathrm{IS}}=\frac{\pi }{8}(\gamma _{h}-\gamma _{l})\left[ 
\frac{1}{(\beta _{h}\gamma _{h})^{2}}-\frac{1}{(\beta _{l}\gamma _{l})^{2}}%
\right] .  \label{28.8}
\end{equation}%
In order to better compare these results, these are all listed in Table IV.

Before concluding this section, we would like to mention that, our current
QHE model can only be implemented with quantum systems with all energy
levels being discrete (all eigen states being bond states). Besides the
harmonic oscillator and the infinite square well system, we can find other
potentials that satisfy discrete spectral structure requirement to implement
our QHE. However, we cannot deal with quantum systems with continuous
spectral structure, e.g., a particle in a Coulumb potential or a finite
square well. We will extend our current study to quantum systems with
continuous spectral structure in future research.

%\begin{widetext}

%\end{widetext}

\section{MAXWELL'S DEMON AND QUANTUM OTTO ENGINE}

In the above discussions, we give clear definitions of microscopic QCE and
QOE cycles through clarifying the basic quantum thermodynamic processes
(e.g., quantum isochoric process and quantum isothermal process). These
results indicate that the properties, such as the operation efficiency, of
macroscopic (classical) heat engines can be obtained from the microscopic
(quantum) level as long as we clarify the basic thermodynamic processes
microscopically. In the previous discussions, our QCE and QOE model show no
contradiction with the thermodynamic laws.

In one of our recent studies \cite{quan4}, we proposed a Maxwell's demon
assisted quantum thermodynamic cycle to study the function of a Maxwell's
demon, and we also studied how it affects the second law of thermodynamics.
It is interesting that when the restoration of the demon are properly
included into the QHE cycle, the efficiency of the Maxwell's demon assisted
QHE cycle has the same form as that for a QOE derived previously (\ref{17}).
Hence, the apparent violation of the second law due to Maxwell's demon is
prevented. In this section, we also would like to study the intrinsic
relation between these two kinds of quantum thermodynamic cycles. We would
also like to add some details about the Maxwell's demon-assisted quantum
thermodynamic cycle proposed in Ref. \cite{quan4} to better demonstrate our
main idea.

\subsection{Maxwell's demon erasure not included in the thermodynamic cycle}

We first analyze a single-reservoir thermodynamic cycle with external
control based on the effective temperature defined above. It can be proved
that the property of this cycle is similar to that of the QHE cycle proposed
in Ref. \cite{quan4}. Our thermodynamic cycle consists of three steps: (1)
quantum rotation, (2) decoherence, and (3) thermalization.

After the thermalization process, the state of the two-level system (with
the ground state $|0\rangle $, the excited state $|1\rangle $ and the level
spacing $\Delta $) can be described by a density matrix 
\begin{equation}
\rho (0)=P_{1}|1\rangle \langle 1|+P_{0}|0\rangle \langle 0|,  \label{50.1}
\end{equation}%
where the probability distributions $P_{1}$ and $P_{0}$ in the two-level
system are determined by the temperature $T_{l}$ of the heat reservoir and \
the level spacing $\Delta $. Then the two-level system, driven by an
external field (we do not treat it as a part of the system for the moment),
undergoes a rotation,

\begin{eqnarray}
|1\rangle &\rightarrow &|\tilde{1}\rangle =\cos \theta |1\rangle +\sin
\theta |0\rangle ,  \label{50.2} \\
|0\rangle &\rightarrow &|\tilde{0}\rangle =-\sin \theta |1\rangle +\cos
\theta |0\rangle .  \notag
\end{eqnarray}%
If the time interval of this rotation is much shorter than the relevant time
of the thermalization, the state $\rho (0)$\ of the two-level system becomes%
\begin{equation}
\rho (1)=P_{1}^{\prime }|1\rangle \langle 1|+P_{0}^{\prime }|0\rangle
\langle 0|+\mathrm{ODT,}
\end{equation}%
where 
\begin{eqnarray}
P_{1}^{\prime } &=&P_{1}\cos ^{2}\theta +P_{0}\sin ^{2}\theta , \\
P_{0}^{\prime } &=&P_{1}\sin ^{2}\theta +P_{0}\cos ^{2}\theta ,  \notag
\end{eqnarray}%
and $\mathrm{ODT}$ denotes the off-diagonal terms, which disappear rapidly
(pure dephasing) due to the coupling between the two-level system and the
reservoir. Actually two effects (dephasing and dissipation) occur when the
two-level system is coupled to the reservoir \cite{sun}. The first process
occurs much faster than the second one, and thus we can consider the two
effects separately. After considering dephasing (but before considering
dissipation), the state $\rho (1)$ of the two-level system becomes%
\begin{equation}
\rho (2)=P_{1}^{\prime }|1\rangle \langle 1|+P_{0}^{\prime }|0\rangle
\langle 0|.  \label{50.4}
\end{equation}%
This state is obviously not in equilibrium with respect to the reservoir at
temperature $T_{l}$. But we can imagine there is such a reservoir at
temperature $T_{h}$, which can be expressed as 
\begin{equation}
T_{h}(P_{0},P_{1},\theta )=\frac{\Delta }{k_{B}}\ln ^{-1}\left( \frac{%
P_{1}\sin ^{2}\theta +P_{0}\cos ^{2}\theta }{P_{1}\cos ^{2}\theta +P_{0}\sin
^{2}\theta }\right) .  \label{50.5}
\end{equation}%
This effective temperature $T_{h}(P_{0},P_{1},\theta )$ possesses some
exotic features. For example, when $\theta =\pi /2$, $P_{1}^{\prime }=P_{0}$
and $P_{0}^{\prime }=P_{1}$. This means that $T_{h}=-T_{l}$ is a
\textquotedblleft negative temperature" since there exists a population
inversion $P_{1}^{\prime }\geq P_{0}^{\prime }$ . Only when $P_{0}^{\prime
}>P_{1}^{\prime }$, $T_{h}$ is positive. Finally, the two-level system is
put into contact with the heat bath for a sufficiently long time. After the
thermalization process, the state $\rho (2)$ returns to $\rho (0)$, and a
thermodynamic cycle is finished and the two-level system seems to extract
work from a single heat bath. We can imagine it as\ a thermodynamic cycle
between two reservoirs with the temperature $T_{l}$ and a virtual
temperature $T_{h}$. Actually contradictions to the second law can appear
due to this \textquotedblleft negative temperature".

It is not surprising that the above result (a contradiction to the second
law) appears since we do not include the controller for the rotation
operators shown in Eq.~(\ref{50.2}). This result is very similar to those
single-particle heat engines assisted by classical Maxwell's demon proposed
by Szilard.

%Figure 7
%\begin{figure}[h]
\begin{figure}[tbp]
\begin{center}
\includegraphics[bb=96 402 500 650, clip, width=8cm]{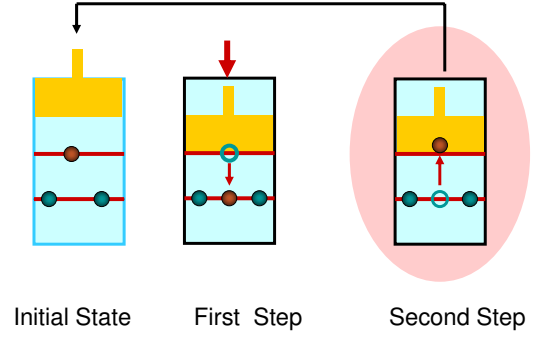}
\end{center}
\caption{(Color online) Schematic diagram of the two-level version of the
Szilard single-particle heat engine. An invisible demon detects the state of
the two level working substance and then controls it to do work: when the
system is in the excited state, the demon makes the working substance flip
to the ground state (the first step); when the system is in the ground
state, the demon does nothing for the working substance. After these
operations the working substance is brought in contact with a very large
heat bath and then thermalized into its initial state (the second step). }
\end{figure}

Now, let us describe a new version of Szilard single-particle heat engine
(see Fig. 8). The working substance of the QHE is a spin (two-level system)
with ground state $|0\rangle $, and excited state $|1\rangle $. The level
spacing is $\Delta $. We first put the two-level system into a heat bath at
temperature $T$. After a sufficiently long thermalization process, the spin
reaches a thermal equilibrium state, which can be described by $\rho (0)$,
similarly defined as above. Second, a demon performs a measurement. If the
measurement result is confirmed that the system is in its upper state, then
the spin is flipped and positive work is done by the spin with an amount $%
\Delta $. If the measurement result confirms that the system is in its lower
state, then no work is done. Then the system is put into contact with the
heat bath and a new cycle starts. This is a two-step Maxwell's
demon-assisted QHE. A similar discussion has been given in Ref. \cite{kieu1}%
. It is easy to see that the net effect of this QHE is to absorb heat from a
single heat bath and convert it into work. On average, the net work done per
cycle is $P_{1}\Delta $. This is a perpetual machine of the second kind.
This apparent violation of the second law is seen because the erasure of the
demon is not included into the QHE cycle.

\subsection{Maxwell's demon erasure included in the thermodynamic cycle}

In Ref. \cite{quan4}, we have demonstrated that, when the erasure of the
information stored in the Maxwell's demon is considered into the QHE cycle,
the apparent violation of the second law does not hold, i.e., there is no
violation of the second law even in the existence of such a Maxwell's demon.

%Figure 8
%\begin{figure}[h]
\begin{figure}[tbp]
\begin{center}
\includegraphics[bb= 81 387 558 685,width=8cm, clip]{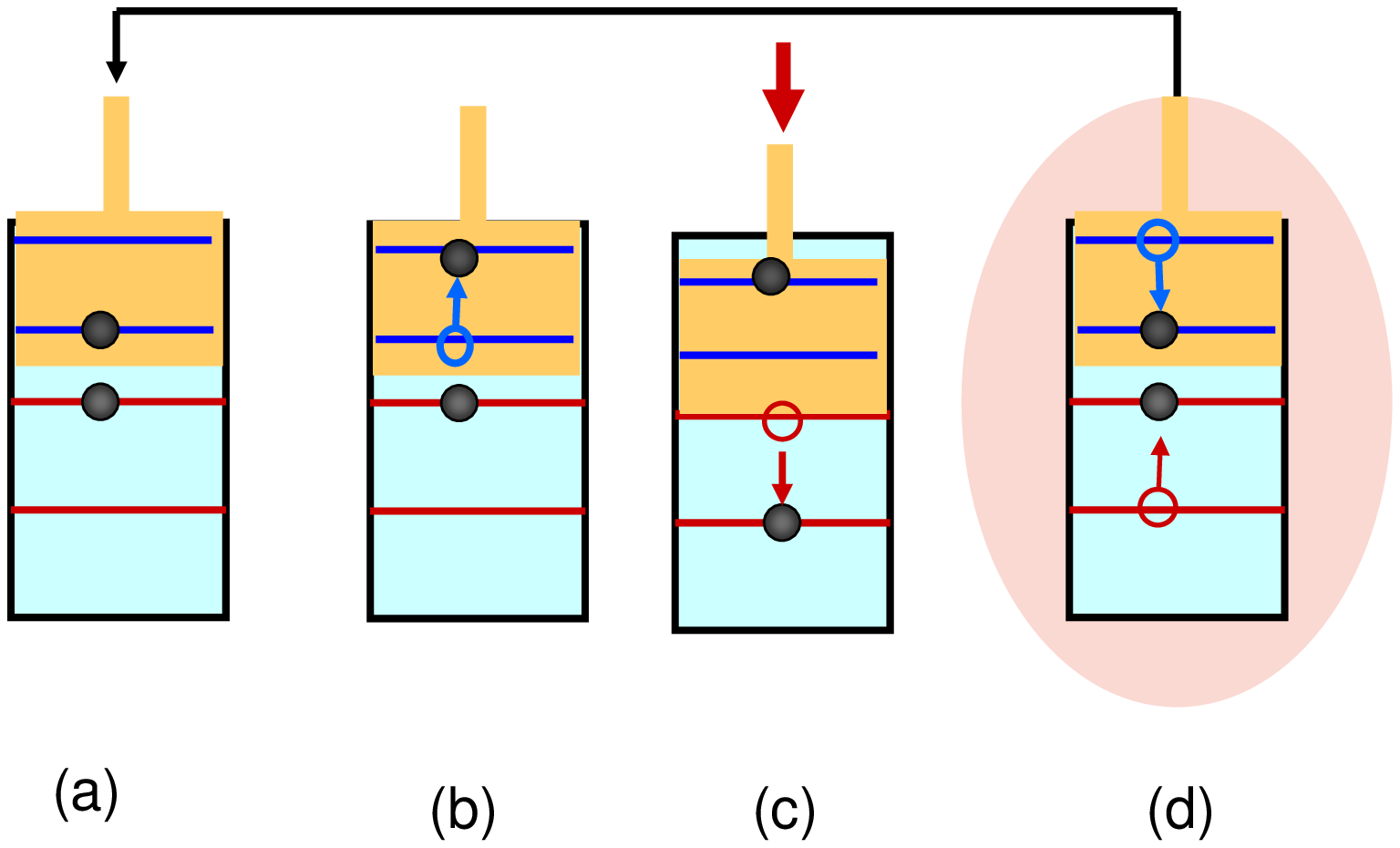}
\end{center}
\caption{(Color online) Schematic illustration for the
Maxwell's-demon-assisted quantum heat engine with four steps: (a): Both the
QHE of the two-level system and a demon is hidden in the \textquotedblleft\
piston" as another two-level system are initialized in thermal equilibrium
states with different temperatures; (b) The demon makes a quantum
non-demolition measurement about the state of the two-level system S with a
CNOT operation: when seeing S in the excited state it records this
information through a transition from its ground state to the excited state;
when S is in its ground state, the demon remains in its original state; (c):
According to the information recorded by the demon, the demon can enable the
system to operate through another CNOT logical gate: If the demon was
encoded in the excited state it will make the system flip; otherwise it
enables the system to remain; (d) both the system and the demon are brought
into contact with their own heat baths and then thermalized with different
temperatures from their own initial states, respectively. During this
process, the information stored in the demon is totally erased and the
entire system completes a quantum thermodynamic cycle. }
\end{figure}

To show the above observation, let us consider in more detail a
thermodynamic cycle including the Maxwell's demon (see the Fig. 9). Let us
explain each step of the QHE cycle proposed in Ref. \cite{quan4}. First, two
qubits (two-level systems) $S$ and $D$ are decoupled and separately coupled
to two heat baths with different temperatures $T_{S}$ and $T_{D}$. After a
period of time longer than both the dephasing time $T_{2}$ and the
relaxation time $T_{1}$, they are thermalized to two equilibrium states $%
\rho _{S}(1)$ and $\ \rho _{D}(1)$%
\begin{equation}
\rho _{F}(1)=P_{F}^{1}|1\rangle \langle 1|+P_{F}^{0}|0\rangle \langle 0|,
\end{equation}%
for $F=S$ and $D$ respectively. The joint thermalized state $\rho (1)=\rho
_{S}(1)\otimes \rho _{D}(1)$ of the total system with $S$ plus $D$ can be
written as%
\begin{eqnarray}
\rho (1) &=&P_{S,D}^{1,1}\,|1,1\rangle \langle
1,1|+P_{S,D}^{1,0}\,|1,0\rangle \langle 1,0|  \label{51} \\
&&+P_{S,D}^{0,1}\,|0,1\rangle \langle 0,1|+P_{S,D}^{0,0}\,|0,0\rangle
\langle 0,0|,  \notag
\end{eqnarray}%
where we have defined the direct product of the eigen state of the two qubits%
\begin{equation}
|q,q^{\prime }\rangle \equiv |q\rangle _{S}\otimes |q^{\prime }\rangle _{D}%
\text{,}\ \ (q,\,q^{\prime }=0,1)  \label{51.2}
\end{equation}%
and the joint probabilities $P_{S,D}^{q,q^{\prime }}\equiv
P_{S}^{q}\,P_{D^{\prime }}^{q^{\prime }}$.

Second, two consecutive unitary operations: a CNOT operation flipping the
demon states only when the working substance system is in its excited state~%
\cite{lloyd}, and the demon controls the system to do work. Physically, the
system experiences a conditional evolution (CEV). The changes of the states
in the two operations can be expressed as follows%
\begin{equation}
\begin{array}{ccccc}
|0,0\rangle & \underrightarrow{\mathrm{CNOT}} & |0,0\rangle & 
\underrightarrow{\mathrm{CEV}} & |0,0\rangle \\ 
|0,1\rangle & \underrightarrow{\mathrm{CNOT}} & |0,1\rangle & 
\underrightarrow{\mathrm{CEV}} & |\tilde{0},1\rangle \\ 
|1,0\rangle & \underrightarrow{\mathrm{CNOT}} & |1,1\rangle & 
\underrightarrow{\mathrm{CEV}} & |\tilde{1},1\rangle \\ 
|1,1\rangle & \underrightarrow{\mathrm{CNOT}} & |1,0\rangle & 
\underrightarrow{\mathrm{CEV}} & |1,0\rangle%
\end{array}%
,  \label{53.1}
\end{equation}%
where $|\tilde{0}\rangle $ and $|1\rangle $ are defined in Eq.~(\ref{50.2}).
These two sub-processes can be realized with two quantum non-demolition
Hamiltonians 
\begin{equation}
H_{\mathrm{CNOT}}=H_{S\rightarrow D},\qquad H_{\mathrm{CEV}}=H_{D\rightarrow
S},
\end{equation}%
where 
\begin{equation}
H_{A\rightarrow B}=\frac{g}{4}(1+\sigma _{Z}^{A})\otimes \sigma _{X}^{B}
\end{equation}%
is a typical CEV Hamiltonian with $A$ controlling $B$. It produces a time
evolution of \ the system $B$ as a quantum rotation defined by Eq.~(\ref%
{50.2}) with $\theta =gt$, when the system $A$ is initially prepared in $%
|1\rangle _{A}$. Here, $\sigma _{j}^{F}$, $F=A,B$, $(j=x,y,z)$ are the Pauli
matrices of the two level system $A$ and $B$. Obviously the CNOT operation
on the demon is given by the time evolution produced by $H_{\mathrm{CNOT}}$
with $\theta =\pi /2$ . $H_{\mathrm{CNOT}}$ performs a quantum
non-demolition measurement of the state of the working substance $S.$

After this CNOT operation, the density matrix changes from $\rho (1)$ to 
\begin{eqnarray}
\rho (2) &=&\mathrm{CNOT}\{\rho (1)\}  \label{54.1} \\
&=&P_{S,D}^{1,1}\,|1,0\rangle \langle 1,0|+P_{S,D}^{1,0}\,|1,1\rangle
\langle 1,1|  \notag \\
&&+P_{S,D}^{0,1}\,|0,1\rangle \langle 0,1|+P_{S,D}^{0,0}\,|0,0\rangle
\langle 0,0|.  \notag
\end{eqnarray}%
From Eq.~(\ref{54.1}) we can derive the reduced density matrices $\rho
_{S}(2)$ and $\rho _{D}(2)$ of the working substance\ and the demon%
\begin{equation}
\rho _{S}(2)=\mathrm{Tr}_{D}[\rho (2)]=\rho _{S}(1)  \label{54.2}
\end{equation}%
and%
\begin{eqnarray}
\rho _{D}(2) &=&\mathrm{Tr}_{S}[\rho (2)]  \label{54.3} \\
&=&(p_{S,D}^{1,1}+p_{S,D}^{0,0})|0\rangle \langle
0|+(p_{S,D}^{1,0}\,+p_{S,D}^{0,1})|1\rangle \langle 1|\ \   \notag
\end{eqnarray}%
respectively. During this process, the system $S$ remains in its initial
state while the demon acquires information. Meanwhile, the internal energy
changes\ at the expense of using an amount of work $W_{D}$ 
\begin{eqnarray}
W_{D} &=&\mathrm{Tr}[H\rho _{D}(2)]-\mathrm{Tr}[H\rho _{D}(1)]  \label{54.4}
\\
&=&\Delta _{D}(P_{D}^{1}-P_{S,D}^{1,0}-P_{S,D}^{0,1}).  \notag
\end{eqnarray}%
From Eqs.~(\ref{51}) and (\ref{54.1}) we see that the total entropy of $S$
and $D$ does not change during the CNOT operation%
\begin{eqnarray}
S(2) &=&-k_{B}[P_{S,D}^{1,1}\ln P_{S,D}^{1,1}+P_{S,D}^{1,0}\ln P_{S,D}^{1,0}
\label{54.5} \\
&&+P_{S,D}^{0,1}\ln P_{S,D}^{0,1}+P_{S,D}^{0,0}\ln P_{S,D}^{0,0}]  \notag \\
&=&-k_{B}[P_{S}^{1}\ln P_{S}^{1}+P_{S}^{0}\ln P_{S}^{0}]  \notag \\
&&-k_{B}[P_{D}^{1}\ln P_{D}^{1}+P_{D}^{0}\ln P_{D}^{0}]  \notag \\
&\equiv &S_{S}(1)+S_{D}(1)\equiv S(1).  \notag
\end{eqnarray}%
Hence, our model verifies the prediction in Refs. \cite{bennett,landauer}
that a measurement do not necessarily lead to entropy increase. Also, we
know from Eq.~(\ref{54.2}) that the entropy of $S$ during the CNOT operation
does not change%
\begin{equation}
S_{S}(2)=S_{S}(1).  \label{54.6}
\end{equation}%
But the entropy of $D$ changes as%
\begin{eqnarray}
\delta S_{D} &=&S_{D}(2)-S_{D}(1)  \label{54.7} \\
&=&-k_{B}[(P_{S,D}^{1,1}+P_{S,D}^{0,0})\ln (P_{S,D}^{1,1}+P_{S,D}^{0,0}) 
\notag \\
&&+(P_{S,D}^{1,0}\,+P_{S,D}^{0,1})\ln (P_{S,D}^{1,0}\,+P_{S,D}^{0,1})] 
\notag \\
&&+k_{B}[P_{D}^{1}\ln P_{D}^{1}+P_{D}^{0}\ln P_{D}^{0}].  \notag
\end{eqnarray}%
We would like to point out that the mutual entropy 
\begin{eqnarray}
S_{M}(2) &=&S_{D}(2)+S_{S}(2)-S(2)  \label{54.8} \\
&=&S_{D}(2)-S_{D}(1).  \notag
\end{eqnarray}%
does not vanish (it vanishes before this CNOT operation). And this
non-vanishing mutual entropy can be used to measure the information acquired
by the demon about the system \cite{zurek}.

Next we consider the changes of entropy and energy after the quantum control
(the CEV) process. The state of the demon and the working system after the
CEV is%
\begin{eqnarray}
\rho (3) &=&\mathrm{CEV}\{\rho (2)\}  \label{53} \\
&=&P_{S,D}^{1,1}\,\left\vert 1,0\right\rangle \left\langle 1,0\right\vert
+P_{S,D}^{1,0}\,\left\vert \tilde{1},1\right\rangle \left\langle \tilde{1}%
,1\right\vert \\
&&+P_{S,D}^{0,1}\,\left\vert \tilde{0},1\right\rangle \left\langle \tilde{0}%
,1\right\vert +P_{S,D}^{0,0}\,\left\vert 0,0\right\rangle \left\langle
0,0\right\vert .  \notag
\end{eqnarray}%
The reduced density matrices of the working substance $\rho _{S}(3)$ and the
demon $\rho _{D}(3)$ are respectively%
\begin{eqnarray}
\rho _{S}(3) &=&\mathrm{Tr}_{D}[\rho (3)]  \label{54.9} \\
&=&P_{S,D}^{1,0}\,\left\vert \tilde{1}\right\rangle \left\langle \tilde{1}%
\right\vert +P_{S,D}^{0,1}\,\left\vert \tilde{0}\right\rangle \left\langle 
\tilde{0}\right\vert  \notag \\
&&+P_{S,D}^{1,1}\,\left\vert 1\right\rangle \left\langle 1\right\vert
+P_{S,D}^{0,0}\,\left\vert 0\right\rangle \left\langle 0\right\vert ,  \notag
\end{eqnarray}%
and%
\begin{equation}
\rho _{D}(3)=\mathrm{Tr}_{S}[\rho (3)]=\rho _{D}(2).  \label{55.0}
\end{equation}%
Similarly, the energy change (work performed by the working system) during
this process is: 
\begin{eqnarray}
W_{S} &=&\mathrm{Tr}[H\rho _{S}(2)]-\mathrm{Tr}[H\rho _{S}(3)]  \label{55.1}
\\
&=&\Delta _{S}(P_{S}^{1}-P_{S,D}^{1,1}-  \notag \\
&&P_{S,D}^{1,0}\left\vert \left\langle \tilde{1}\right. \left\vert
1\right\rangle \right\vert ^{2}-P_{S,D}^{0,1}\left\vert \left\langle \tilde{0%
}\right. \left\vert 1\right\rangle \right\vert ^{2}).  \notag
\end{eqnarray}%
In particular, when we choose $\theta =\pi /2$, the CEV is a CNOT, and the
reduced density matrix of $S$ can be written as%
\begin{equation}
\rho _{S}(3)=P_{D}^{1}\,\left\vert 1\right\rangle \left\langle 1\right\vert
+P_{D}^{0}\,\left\vert 0\right\rangle \left\langle 0\right\vert ,
\label{55.2}
\end{equation}%
and the entropy of the system is 
\begin{equation*}
S_{S}(3)=-k_{B}[P_{D}^{1}\,\ln P_{D}^{1}\,+P_{D}^{0}\ln P_{D}^{0}]\equiv
S_{D}(1).
\end{equation*}%
So the change of the entropy of the working substance in the CEV process is%
\begin{equation}
\delta S_{S}=S_{S}(3)-S_{S}(2)=S_{D}(1)-S_{S}(1),
\end{equation}%
where we have used Eq.~(\ref{54.6}). But from Eqs.~(\ref{54.5}), (\ref{53})
and (\ref{55.0}), we know that both the total entropy of $S$ plus $D$ and
the entropy of the demon $D$ do not change, i.e.%
\begin{eqnarray}
S(1) &=&S(2)=S(3),  \label{55.3} \\
S_{D}(3) &=&S_{D}(2).  \notag
\end{eqnarray}

Finally, $S$ plus $D$ are decoupled and put into contact with their own
baths, and a new cycle starts. In the thermalizaiton process, no work is
done, but heat is exchanged between the heat baths and $S$ and $D$. The
thermalization process is an information-erasure process. This kind of
zero-work erasure with a low temperature reservoir was first introduced in
Ref. \cite{lubkin}, and studied afterwards in Ref. \cite{plenio}. In the
thermalization process the heat absorbed by $S$ is%
\begin{eqnarray}
Q_{\mathrm{in}} &=&\mathrm{Tr}[H\rho _{S}(1)]-\mathrm{Tr}[H\rho _{S}(3)] \\
&=&\Delta _{S}(P_{S,D}^{1,0}-P_{S,D}^{1,0}\left\vert \left\langle \tilde{1}%
\right. \left\vert 1\right\rangle \right\vert ^{2}-P_{S,D}^{0,1}\left\vert
\left\langle \tilde{0}\right. \left\vert 1\right\rangle \right\vert ^{2}), 
\notag
\end{eqnarray}%
and the heat released by the demon $D$ is%
\begin{eqnarray}
Q_{\mathrm{out}} &=&\mathrm{Tr}[H\rho _{D}(3)]-\mathrm{Tr}[H\rho _{D}(1)]
\label{55.5} \\
&=&\Delta _{D}\left( P_{S,D}^{1,0}\,+P_{S,D}^{0,1}-P_{D}^{1}\right) .  \notag
\end{eqnarray}

Now we include the erasure of the memory of Maxwell's demon into the
thermodynamic cycle, and we will show that, under certain conditions, our
composite QHE is equivalent to a simple QOE.

For each cycle described above, we are now able to calculate the work $W$
performed by the heat engine%
\begin{eqnarray}
W &=&W_{S}-W_{D}  \label{53.2} \\
&=&\Delta _{S}(P_{S,D}^{1,0}-P_{S,D}^{1,0}\left\vert \left\langle \tilde{1}%
\right. \left\vert 1\right\rangle \right\vert ^{2}  \notag \\
&&-P_{S,D}^{0,1}\left\vert \left\langle \tilde{0}\right. \left\vert
1\right\rangle \right\vert ^{2})-\Delta _{D}\left(
P_{S,D}^{1,0}-P_{S,D}^{1,1}\right) .  \notag
\end{eqnarray}%
It can be checked that in the thermodynamic cycle $W=Q_{\mathrm{in}}-Q_{%
\mathrm{out}}$. This is just the first law of thermodynamics. The
positive-work condition can be derived from Eq.~(\ref{53.2}) 
\begin{equation}
T_{S}\geq T_{D}\left( \frac{\Delta _{D}}{\Delta _{S}}\right) .  \label{57.1}
\end{equation}%
Notice that when we choose the CEV to be the special case $\theta =\pi /2$,
(i.e., a CNOT) the heat absorbed $Q_{\mathrm{in}}$ and positive work $W$
done during a cycle can be simplified to%
\begin{eqnarray}
Q_{\mathrm{in}} &=&\Delta _{S}\left( P_{S,D}^{1,0}-P_{S,D}^{0,1}\right) ,
\label{53.4} \\
W &=&\Delta _{S}\left( P_{S,D}^{1,0}-P_{S,D}^{0,1}\right) -\Delta _{D}\left(
P_{S,D}^{1,0}-P_{S,D}^{1,1}\right) ,  \label{53.5}
\end{eqnarray}%
and 
\begin{equation}
\eta =\frac{W}{Q_{\mathrm{in}}}=1-\frac{\Delta _{D}}{\Delta _{S}}\frac{%
\left( P_{S,D}^{1,1}-P_{S,D}^{1,0}\right) }{\left(
P_{S,D}^{1,0}-P_{S,D}^{0,1}\right) }.  \label{53.7}
\end{equation}%
If we further assume the temperature $T_{D}$ to be so low that $\exp (-\beta
_{D}\Delta _{D})\ll 1$, i.e., the demon is \textquotedblleft erased" nearly
to its ground state $\rho _{D}\left( 1\right) \approx \left\vert
0_{D}\right\rangle \langle 0_{D}|$ (The zero-entropy \textquotedblleft
standard state" was also discussed in Ref. \cite{bennett,lubkin,plenio}).
Then the efficiency of our QHE, Eq.~(\ref{53.7}), becomes%
\begin{equation}
\eta =1-\frac{\Delta _{D}}{\Delta _{S}}.  \label{57.2}
\end{equation}%
This is the efficiency of a simple QOE cycle without Maxwell's demon, as
shown in Eq.~(\ref{17}). Otherwise the operation efficiency in Eq.~(\ref%
{53.7}) is less than the efficiency of a simple QOE cycle. This is because
i) among all CEVs, the CNOT is the optimal operation to extract work, and
ii) when $T_{D}$ is vanishingly small, the demon can be restored to a
zero-entropy \textquotedblleft standard state" to acquire information about
the system in the most efficient way.

\section{CONCLUSIONS AND REMARKS}

By defining the quantum version of basic thermodynamic processes, we study
the basic properties of a QCE. To construct a QCE cycle, two preconditions
about the quantum adiabatic process are required: first, all energy gaps of
the working substance must change by the same ratio, such as in harmonic
oscillators and infinite square well potentials. Second, the change of the
ratio of the energy gaps must equal the ratio of the temperatures of the two
heat baths. We find that the working efficiency for the QCE is the same as
that of the classical Carnot engine though the internal energy may change in
the quantum isothermal process. We also study the properties of the QOE and
compare these with the classical Otto engine, and we find that the
efficiency and positive-work condition are the same when expressed in terms
of temperatures (see Table III). Through comparing the thermodynamic cycles
of the two QHEs, we clarify the relationship between them, and we
demonstrate that the QCE (QOE) cycle can be modeled as an infinite number of
small QOE (QCE) cycles. We also discuss some experimentally realizable
physical systems that can be used to implement our QHE. Finally, through a
generalized QOE, we demonstrate that there is no violation of the second
law, even when there is a Maxwell's demon. This is a prediction of
Landauer's principle~\cite{landauer,bennett2003,demon,maruyama, scully2005}.

Before concluding this paper, we want to emphasize three points. First, in
our present study the working substance is always assumed to be in a thermal
equilibrium state, and quantum coherence is not considered here, our study
is not related with the result that the efficiency of a QHE is less than its
classical counterpart $\eta _{\mathrm{Q}}\leq \eta _{\mathrm{C}}$, as in
Ref.~\cite{lloyd}. Second, we only consider quasi-static process. Hence, the
time intervals of these processes are infinitely long and the output power
is vanishingly small. Recently, finite-time QHE cycles~\cite{kosloff1,curzon}
(nonzero output power) and friction-like behavior~\cite{kosloff2} of the QOE
were studied, where the increase of power occurs at the expense of
decreasing operation efficiency. Third, we will further extend our current
study to QHE with quantum-many-body system as the working substance. In this
extened case, we will consider the quantum statistical effects, e.g., the
Bose-Einsten condensation, of the working substance. We believe these
quantum effects will advance our understanding of the relation between
quantum mechanics and thermodynamics and bring important insights into some
fundamentle problems in quantum thermodynamics.

\section{acknowledgments}

We thank Dr. K. Maruyama for helpful discussions. FN was supported in part
by the US National Security Agency (NSA), Army Research Office (ARO),
Laboratory of Physical Sciences (LPS), and the National Science Foundation
grant No.~EIA-0130383. CPS was supported in part by the NSFC with grant Nos.
90203018, 10474104 and 60433050; and the National Fundamental Research
Program of China with Nos. 2001CB309310 and 2005CB724508.

\begin{appendix}

\setcounter{section}{0} \setcounter{equation}{0} \renewcommand{\thesection}{%
\Alph{section}}

\appendix
\section{INVARIANCE UNDER ENERGY SHIFT}

The amount of positive work $W_{\mathrm{O}}$ done during a QOE cycle
remains invariant under the uniform shift of all energy levels. If
we shift all the
energy levels $E_{n}^{l}$ in Fig.1%
\begin{equation}
\tilde{E}_{n}^{l}=E_{n}^{l}-\delta ,  \label{30}
\end{equation}%
the amount of work done during a QOE cycle becomes%
\begin{equation*}
\tilde{W}_{\mathrm{O}}=\sum_{n}\,(P_{n}^{h}-\tilde{P}_{n}^{l})(E_{n}^{h}-%
\tilde{E}_{n}^{l}),
\end{equation*}%
where $\tilde{P}_{n}^{l}$ are the occupation probabilities after the
energy
levels are shifted. It is easy to find that the occupation probabilities $%
\tilde{P}_{n}^{l}$ remain invariant under such an energy shift%
\begin{eqnarray}
\tilde{P}_{n}^{l} &=&e^{-(E_{n}^{l}-\delta )\beta _{l}}\left[
\sum_{n}e^{-(E_{n}^{l}-\delta )\beta _{l}}\right] ^{-1}  \label{31} \\
&=&e^{-E_{n}^{l}\beta _{l}}\left[ \sum_{n}e^{-E_{n}^{l}\beta
_{l}}\right] ^{-1}=P_{n}^{l}.  \notag
\end{eqnarray}%
Thus, after the energy levels shift, the net work
$\tilde{W}_{\mathrm{O}}$ done during a QOE cycle can be simplified
to

\begin{eqnarray}
\tilde{W}_{\mathrm{O}}
&=&\sum_{n}\,(P_{n}^{h}-P_{n}^{l})(E_{n}^{h}-E_{n}^{l}+\delta )
\label{32}
\\
&=&\sum_{n}\,(P_{n}^{h}-P_{n}^{l})(E_{n}^{h}-E_{n}^{l}),  \notag
\end{eqnarray}%
which is just the net work $W_{\mathrm{O}}$ done by the working
system
during a QOE cycle before the level shift. Similarly we can prove that $W_{%
\mathrm{O}}$ remains invariant under the shift $\tilde{E}_{n}^{h}=E_{n}^{h}-%
\delta $.

Following the same way we can prove that other properties of the
QHE, such as operation efficiency and positive work conditions, are
invariant under uniform shift of all energy levels as well. Hence,
we can simply assume the eigen energy of the ground state to be zero
$E_{0}^{h}=E_{0}^{l}=0$. This energy shift is convenient for our
discussion about QHE later. This result can be easily generalized to
any QCE cycle, because in Sec. V, we demonstrate that a QCE can be
modelled as an infinite number of small QOE cycles.

\section{INTERNAL ENERGY OF THE WORKING SUBSTANCE IN QUANTUM ISOTHERMAL PROCESSES}

The internal energy of a system is%
\begin{equation}
U(i)=\mathrm{Tr}[\rho (i)H(i)].  \label{33}
\end{equation}%
In an isothermal expansion process, all energy levels change in the
same ratio
\begin{equation*}
E_{n}\rightarrow \zeta E_{n},\ \ \ \ n=0,1,2,\cdots ;
\end{equation*}%
where $\zeta $ is the ratio of energy levels and $0<\zeta <1$. When
the
energy levels change, the internal energy of the system can be rewritten as%
\begin{equation}
U(\zeta )=\sum_{n}\frac{\zeta E_{n}}{Z(\zeta )}\exp [-\beta
_{h}\zeta E_{n}], \label{34}
\end{equation}%
where%
\begin{equation}
Z(\zeta )=\sum_{n}\exp [-\beta _{h}\zeta E_{n}]  \label{35}
\end{equation}%
is the partition function. To test whether the internal energy of
the system
is invariant under the change of energy levels, we take the derivative%
\begin{eqnarray}
\frac{dU(\zeta )}{d\zeta } &=&\sum_{n}\frac{E_{n}}{Z(\zeta
)}(1-\zeta \beta
_{h}E_{n})\exp [-\beta _{h}\zeta E_{n}]  \label{36} \\
&&+\zeta \beta _{h}\left[ \sum_{n}\frac{E_{n}}{Z(\zeta )}\exp
[-\beta _{h}\zeta E_{n}]\right] ^{2}.  \notag
\end{eqnarray}

For two-level systems, there is only one term in the sum over $n$,
because
we assume the eigen energy of the ground state to be zero. Thus Eq.~(\ref{36}%
) can be simplified as%
\begin{equation}
\frac{dU(\zeta )}{d\zeta }=\frac{E_{e}}{Z(\zeta )}\exp [-\beta
_{h}\zeta E_{e}]\left[ 1-\frac{\zeta \beta _{h}E_{n}}{Z(\zeta
)}\right] ,  \label{37}
\end{equation}%
where $E_{e}$ is the eigen energy of the excited state of the
two-level system. The rhs of Eq.~(\ref{37}) is obviously nonzero.

For a harmonic oscillator with the eigenfrequency $\omega $, we
assume the eigen energy of the ground state to be zero (neglecting
the ground state
energy $\hbar \omega /2$), then Eq.~(\ref{36}) can be simplified to%
\begin{eqnarray}
\frac{dU(\zeta )}{d\zeta } &=&-\left[ \frac{\hbar \omega (\beta
_{h}\zeta \hbar \omega +1)}{\exp [\beta _{h}\zeta \hbar \omega
]-1}\right]   \label{38}
\\
&&-\left[ \frac{\hbar \omega \beta _{h}\zeta }{\exp [\beta _{h}\zeta
\hbar \omega ]-1}\right] ^{2}.  \notag
\end{eqnarray}%
The rhs of Eq.~(\ref{38}) is obviously nonzero.

For a particle confined in an infinite square potential, $E_{n}=\gamma n^{2}$%
, where $\gamma $ has been defined in Eq.~(\ref{29.7}). Then
Eq.~(\ref{36})
can be simplified to%
\begin{eqnarray}
\frac{dU(\zeta )}{d\zeta } &=&\sum_{n}\frac{\gamma n^{2}-\beta
_{h}\zeta \gamma ^{2}n^{4}}{Z(\zeta )}\exp [-\beta _{h}\zeta \gamma
n^{2}]  \label{39}
\\
&&+\beta _{h}\zeta \left[ \sum_{n}\frac{\gamma n^{2}}{Z(\zeta )}\exp
[-\beta _{h}\zeta \gamma n^{2}]\right] ^{2},  \notag
\end{eqnarray}%
where%
\begin{equation}
Z(\zeta )=\sum_{n}\exp [-\beta _{h}\zeta \gamma n^{2}]  \label{40}
\end{equation}%
is the partition function. We make an approximation%
\begin{equation}
Z(\zeta )\approx \int_{0}^{\infty }\exp [-\beta _{h}\zeta \gamma n^{2}]dt=%
\frac{1}{2}\sqrt{\frac{\pi }{\beta _{h}\zeta \gamma }},  \label{41}
\end{equation}%
and then Eq.~(\ref{39}) can be simplified to%
\begin{equation}
\frac{dU(\zeta )}{d\zeta }\ =\ -\frac{1}{\beta _{h}\zeta }.
\label{42}
\end{equation}%
The rhs of Eq.~(\ref{42}) is nonzero, either. Hence, we conclude
that in an isothermal process, the derivation of the internal energy
over the energy levels in Eq.~(\ref{36}) is always nonzero.
Accordingly, the internal energy of the system varies with the
change of the energy levels.

\section{THERMODYNAMIC REVERSIBILITY OF AN INFINITE NUMBER OF INFINITESIMAL
QUANTUM ISOTHERMAL PROCESSES}

We consider a two-level system interacting with a heat bath. The
temperature of the heat bath is well controlled so that the
two-level system is always in thermal equilibrium with the heat
bath. Now let us calculate the entropy (Von Neumann entropy)
increase in the two-level system and the entropy (thermodynamic
entropy) decrease in the heat bath. We assume the two-level
system initially in thermal equilibrium with a heat bath at the temperature $%
T_{h}^{1}$ (see Eq. (\ref{28}) and Fig. 6), and finally the
temperature of the heat bath is controlled to increase to $T_{h}$
(see Fig.6). The entropy of the initial state of the system at
$A^{\prime }$ (see Fig. 6) is
\begin{eqnarray}
&&S_{\mathrm{TLS}}(A^{\prime }) \\
&=&-k_{B}(\Lambda _{-}^{\prime }\exp [-\Delta _{h}\beta _{h}^{\prime
}]\ln (\Lambda _{-}^{\prime }\exp [-\Delta _{h}\beta _{h}^{\prime
}])+\Lambda
_{-}^{\prime }\ln \Lambda _{-}^{\prime })  \notag \\
&=&\frac{\Delta _{h}}{T_{h}^{\prime }}\Lambda _{+}^{\prime
}-k_{B}\ln (\Lambda _{+}^{\prime }\exp [\Delta _{h}\beta
_{h}^{\prime }]),  \notag
\end{eqnarray}%
where $\beta _{h}^{\prime }=1/(k_{B}T_{h}^{1})$ and
\begin{eqnarray}
\Lambda _{\pm }^{\prime } &=&\frac{1}{1+\exp [\pm \Delta _{h}\beta
_{h}^{\prime }]},  \notag \\
\Lambda _{\pm } &=&\frac{1}{1+\exp [\pm \Delta _{h}\beta _{h}]}.
\end{eqnarray}%
Similarly, we obtain the entropy of the final state of the two-level
system at $B^{\prime }$ (see Fig. 6) is
\begin{equation}
S_{\mathrm{TLS}}(B^{\prime })=\frac{\Delta _{h}}{T_{h}}\Lambda
_{+}-k_{B}\ln (\Lambda _{+}\exp [\Delta _{h}\beta _{h}]).
\end{equation}%
So the entropy increase of the two-level system in the infinite
number of infinitesimal quantum isothermal processes ($A^{\prime
}\longrightarrow
B^{\prime }$ in Fig. 6) is%
\begin{eqnarray}
dS_{\mathrm{TLS}} &=&S_{\mathrm{TLS}}(B^{\prime })-S_{\mathrm{TLS}%
}(A^{\prime })  \label{c4} \\
&=&\frac{\Delta _{h}}{T_{h}}\Lambda _{+}-k_{B}\ln \left( \Lambda
_{+}\exp
[\Delta _{h}\beta _{h}]\right)   \notag \\
&&-\left[ \frac{\Delta _{h}}{T_{h}^{1}}\Lambda _{+}^{\prime
}-k_{B}\ln
\left( \Lambda _{+}^{\prime }\exp [\Delta _{h}\beta _{h}^{\prime }]\right) %
\right] .  \notag
\end{eqnarray}%
Next we calculate the entropy decrease of the heat bath in the same
process. The thermodynamic entropy decrease of the heat bath due to
its coupling to
the two-level system can be expressed as%
\begin{eqnarray}
dS_{\mathrm{Bath}} &=&\int_{T_{h}^{1}}^{T_{h}}\frac{\Delta
_{h}dp_{e}}{T}
\label{b} \\
&=&k_{B}\int_{T_{h}^{1}}^{T_{h}}\left[ (\Lambda _{+})\left( \frac{\Delta _{h}%
}{k_{B}}\right) ^{2}-(\Lambda _{+})^{2}\left( \frac{\Delta _{h}}{k_{B}}%
\right) ^{2}\right] \frac{dT}{T}. \notag
\end{eqnarray}%
We apply the transformation%
\begin{eqnarray}
T &=&\frac{\Delta _{h}}{k_{B}\ln t}, \\
dT &=&\frac{\Delta _{h}}{k_{B}}\left( -\frac{1}{\ln ^{2}t}\right) \frac{dt}{t%
}.  \notag
\end{eqnarray}%
Under this transformation $dS_{\mathrm{Bath}}$ (\ref{b}) can be
further given as
\begin{eqnarray}
&&dS_{\mathrm{Bath}}  \label{c8} \\
&=&k_{B}\left[ \left. \left( \frac{\ln t}{1+t}\right) \right\vert
_{\exp [\Delta _{h}\beta _{h}^{\prime }]}^{\exp [\Delta _{h}\beta
_{h}]}-\left. \ln \left( \frac{t}{1+t}\right) \right\vert _{\exp
[\Delta _{h}\beta
_{h}^{\prime }]}^{\exp [\Delta _{h}\beta _{h}]}\right]   \notag \\
&=&\frac{\Delta _{h}}{T_{h}}\Lambda _{+}-k_{B}\ln \left( \Lambda
_{+}\exp
[\Delta _{h}\beta _{h}]\right)   \notag \\
&&-\left( \frac{\Delta _{h}}{T_{h}^{\prime }}\Lambda _{+}^{\prime
}-k_{B}\ln \Lambda _{+}^{\prime }\exp [\Delta _{h}\beta _{h}^{\prime
}]\right) .  \notag
\end{eqnarray}%
From Eqs. (\ref{c4}) and (\ref{c8}) we can see that $%
dS_{\mathrm{Bath}}=dS_{\mathrm{TLS}}$, i.e., the entropy decrease in
the bath is equal the entropy increase in the two-level system. Thus
we proved the total entropy conserves in these infinite number of
infinitesimal quantum isothermal processes, and these processes are
thermodynamically reversible.

\end{appendix}


\begin{thebibliography}{99}
\bibitem{explain} From Eq. (45) of Ref. \cite{quan1} and the discussion
therein we can see that results about QHE based on a two-level system can be
generalized to a multi-level case (no matter how the spectral structure is)
straightforwardly as long as all energy levels of the multi-level system
change in the same ratio in the quantum adiabatic process. The harmonic
oscillator and the particle confined in an infinite square well potential
are two good example quantum systems that can be applied to implement the
QHE.
\end{thebibliography}
\end{document}